\documentclass{article}
\usepackage{authblk}
\providecommand{\keywords}[1]{\textbf{\textit{Keywords---}} #1}
\usepackage{graphicx}
\usepackage{url}
\usepackage{hyperref}
\hypersetup{
    colorlinks=true,
    linkcolor=blue,
    filecolor=magenta,      
    urlcolor=blue,
    citecolor=blue,
}
\usepackage{amscd}
\usepackage{multirow}
\usepackage{amssymb}
\usepackage{graphicx}
\usepackage{amsmath,amsthm}
\newtheorem{remark}{Remark}
\usepackage{amscd}
\usepackage{amsfonts}
\usepackage{longtable}% for long tables

\usepackage{booktabs}
\usepackage{natbib}

\usepackage{dsfont}
\usepackage{caption} %pour figure caption in minipage which does not allow envir figure
\usepackage{float} % pour forcer placement figure avec [H]
\usepackage{graphicx}
\usepackage{xifthen}
\usepackage{bbm}
\usepackage{mathtools}
\usepackage{amssymb}
\usepackage{bm}

\usepackage{xcolor}
\usepackage[export]{adjustbox}

\newcommand{\mb}{\mathbf}

\newtheorem{model}{Model}

\newcommand{\ie}{\emph{i.e.}{}}

\newcommand{\eg}{\emph{e.g.}{}}

\def\mb{\mathbf}
\def\point{\,\cdot\,}

\newcommand{\EE}[1][]{%
  \ifthenelse{\isempty{#1}}%
    {\mathbb{E}}% if #1 is empty
    {\mathbb{E}\left(#1\right)}% if #1 is not empty
}
\newcommand{\PP}[1][]{%
  \ifthenelse{\isempty{#1}}%
    {\mathbb{P}}% if #1 is empty
    {\mathbb{P}\left(#1\right)}% if #1 is not empty
  }

\def\P{\PP}
\newcommand{\rset}{\mathbb{R}}

\newcommand{\qset}{\mathbb{Q}}
\newcommand{\ninf}[1]{\| {#1}\|_{\infty}}
\newcommand{\sumnorm}[1]{\| {#1}\|}
\newcommand{\simplex}{\mathcal{S}}
\newcommand{\diri}{\varphi}%%mathcal{D}\mathrm{ir}}
\newcommand{\dexp}{\mathcal{E}\mathit{xp}}%%mathcal{D}\mathrm{ir}}
\newcommand{\dnoise}{f_{\varepsilon}}%%mathcal{D}\mathrm{ir}}
\newcommand{\ud}{\mathrm{d}}
\newcommand{\dd}{\{1,\ldots,d\}}

\newcommand{\nto}{\;\xrightarrow[n \rightarrow \infty]{}\;}  
\newcommand{\given}[1][{}]{\;\middle\vert\;{#1} }

\newcommand{\bzero}{{\mathbf 0}}
\newcommand{\1}{\mathbbm{1}}
\newcommand{\cone}{\mathcal{C}}
\newcommand{\rect}{\mathcal{R}}

\newcommand{\bepsilon}{\boldsymbol{\varepsilon}}
\newcommand{\blambda}{\boldsymbol{\lambda}}
\newcommand{\bnu}{\boldsymbol{\nu}}
\newcommand{\bgamma}{\boldsymbol{\gamma}}
\newcommand{\brho}{\boldsymbol{\rho}}
\newcommand{\bpi}{\boldsymbol{\pi}}
\newcommand{\btheta}{\boldsymbol{\theta}}
\DeclareMathOperator*{\argmax}{arg\,max}
\newcommand{\err}{\mathrm{err}}

\begin{document}
\title{A Multivariate Extreme Value Theory Approach \\to Anomaly Clustering and Visualization}
% \title{Insert your title here%\thanks{Grants or other notes
% %about the article that should go on the front page should be
% %placed here. General acknowledgments should be placed at the end of the article.}
% }
% \subtitle{Do you have a subtitle?\\ If so, write it here}

%\titlerunning{A MEVT Approach to Anomaly Clustering and Visualization}        % if too long for running head

% \author[SC]{Ma\"el Chiapino}
% \author[SC]{St\'ephan Cl\'emen\c{c}on}
% \author[SC]{Anne Sabourin}
% \author[VF]{Vincent Feuillard}
% \address[SC]{LTCI Telecom ParisTech, Universit\'e Paris Saclay}
% \address[VF]{AIRBUS}

\author[1]{Ma\"el Chiapino}         
\author[1]{  Stephan Cl\'emen\c{c}on}
  \author[2]{Vincent Feuillard} 
\author[1]{Anne Sabourin}
%         Second Author %etc.

% %\authorrunning{Short form of author list} % if too long for running head

 \affil[1]{ LTCI, T\'el\'ecom Paris, Institut polytechnique de Paris, France\\
%               Tel.: +123-45-678910\\
%               Fax: +123-45-678910\\
            \texttt{anne.sabourin@telecom-paristech.fr}
%               \emph{Present address:} of F. Author  %  if needed
           }
             \affil[2]{
              Airbus Central R\&T, AI Research
        }

\date{\today}
% The correct dates will be entered by the editor

\maketitle
%\linenumbers

\begin{abstract}
  In a wide variety of situations, anomalies in the behaviour of a
  complex system, whose health is monitored through the observation of
  a random vector $\mb X=(X_1,\; \ldots,\; X_d)$ valued in
  $\mathbb{R}^d$, correspond to the simultaneous occurrence of extreme
  values for certain subgroups $\alpha\subset\{1,\; \ldots,\; d \}$ of
  variables $X_j$.  Under the heavy-tail assumption, which is
  precisely appropriate for modeling these phenomena, statistical
  methods relying on multivariate extreme value theory have been
  developed in the past few years 
  for identifying such
  events/subgroups. 
  This paper exploits this approach much further by means of
  a novel mixture model that permits to describe the distribution of
  extremal observations and where the anomaly type $\alpha$ is viewed
  as a latent variable. One may then take advantage of the model by assigning to
  any  extreme point  a posterior probability % of occurence
  for each anomaly type $\alpha$, defining implicitly a similarity
  measure between anomalies. 
  It is explained at length how the latter 
  permits to cluster extreme observations and 
  obtain an informative planar representation of anomalies using
  standard graph-mining tools. The relevance and usefulness of the clustering and
  $2$-d visual display thus designed is illustrated on simulated datasets and on real observations as well,
  in the aeronautics application domain.

  \keywords{
Anomaly detection, clustering, graph-mining, latent variable analysis, mixture modelling, multivariate extreme value theory, visualization}

\end{abstract}

%\end{frontmatter}

\section{Introduction}\label{sec:intro}
Motivated by a wide variety of applications ranging from fraud
detection to aviation safety management through the health monitoring
of complex networks, data center infrastructure management or food
risk analysis, unsupervised anomaly detection is now the subject of
much attention in the data science literature, see \textit{e.g.}
\cite{GMM12,FP97,VCTWMS}. In frequently encountered practical
situations and from the viewpoint embraced in this paper, anomalies
coincide with rare measurements that are extremes, \textit{i.e.}
located far from central statistics such as the sample mean.  In the
$1$-d setting, numerous statistical techniques for anomaly detection
are based on a parametric representation of the tail of the observed
univariate probability distribution, relying on \emph{extreme value
  theory} (EVT), see \textit{e.g.}
\cite{Clifton2011,Lee2008,Roberts2000,Tressou2008} among others. In
(even moderately) large dimensional situations, the modelling task
becomes much harder. Many nonparametric heuristics for supervised
classification have been adapted, substituting rarity for labeling,
see \textit{e.g.} \cite{Scholkopf2001}, \cite{SHS05} or
\cite{Liu2008}.  In the unsupervised setting, several extensions of
the basic linear Principal Component Analysis for dimensionality reduction and
visualization techniques have been proposed in the statistics and
data-mining literature, accounting for non linearities or increasing
robustness for instance, \textit{cf} \cite{Kegl08} and
\cite{Kriegel08}. 
These approaches intend to describe parsimoniously
the `center' of a massive data distribution, see \textit{e.g.}
\cite{Naikbook} and the references therein. Similarly, for clustering purposes, 
several  multivariate heavy-tailed  distributions have been proposed that are robust to the presence of outliers, see \emph{e.g.} \cite{forbes2014new}, \cite{punzo2018multiple}. 
However the issue of
clustering extremes or outliers is only recently receiving attention,
at the instigation of industrial applications such as those mentioned
above and because of the increasing availability of extreme
observations in databases: generally out-of-sample in the past,
extreme values are becoming observable in the Big Data era.  It is the
goal of the present article to propose a novel mixture model-based
approach for clustering extremes in the multivariate setup,
\textit{i.e.} when the observed random vector
$\mb X=(X_1,\; \ldots,\; X_d)$ takes its values in the positive
orthant of the space $\mathbb{R}^d$ with $d>1$ equipped with the
sum-norm
$\sumnorm{(x_1,\; \ldots,\; x_d)}=\sum_{1\leq j\leq d}\vert x_j\vert$:
'extremes' coinciding then with values $x$ such that
$\P[ \sumnorm{ \mb X} > \sumnorm{\mb x}]$ is 'extremely
small'. Precisely, it relies on a dimensionality reduction technique
of the tail distribution recently introduced in \cite{goix2017sparse}
and \cite{GoixAISTATS16},  and referred to as the DAMEX
algorithm. Based on multivariate extreme value theory (MEV theory),
the latter method may provide a hopefully sparse representation of the
support
of % the measure describing he extremal behaviour of the r.v. $\mb X$.
the angular measure related to the supposedly heavy-tailed
distribution of the random vector $\mb X$. As the angular measure
asymptotically describes the dependence structure of the variables
$X_j$ in the extremal domain (and, roughly speaking, permits to assign
limit probabilities to directions $\mb x/\sumnorm{\mb x}$ in the unit
sphere along which extreme observations may occur), this statistical
procedure identifies the groups $\alpha\subset\{1,\; \ldots,\; d\} $
of feature indices such that the collection of variables
$\{X_j:\; j\in \alpha \}$ may be simultaneously very large, while the
others, the $X_j$'s for $j\notin \alpha$, remain small. Groups of this
type are  in 1-to-1 correspondence with the faces
$\Omega_{\alpha}=\{\mb x\in \mathbb{R}^d:\; \sumnorm{ \mb x}=1,\;
x_j=0 \text{ if } j\notin \alpha \text{ and } x_j>0 \text{ if } j\in
\alpha\}$ of the unit sphere composing the support of the angular
measure. In practice, a sparse representation of the extremal
dependence structure is obtained when only a few such groups of
variables can be exhibited (compared to $2^d-1$) and/or when these
groups involve a small number of variables (with respect to $d$). Here
we develop this framework further, in order to propose a (soft)
clustering technique in the region of extremes and derive effective
$2$-d visual displays, sheding light on the structure of
anomalies/extremes in sparse situations. This is achieved by modelling
the distribution of extremes as a specific \textit{mixture model},
where each component generates a different type $\alpha$ of
extremes. In this respect, the present paper may be seen as an
extension of \cite{boldi2007mixture, sabourin2014bayesian}, where a Bayesian inference framework is designed for moderate
dimensions ($d\le 10$ say) and situations where the sole
group of variables with the potential of being simultaneously large is
$\{1,\ldots, d\}$ itself.  In the context of  mixture modelling (see \textit{e.g.} \cite{FCR18}), the
Expectation-Maximization algorithm (EM in abbreviated form) permits to
partition/cluster the set of extremal data through the statistical
recovery of \textit{latent observations}, as well as posterior
probability distributions (inducing a soft clustering of the data in a
straighforward manner) and, as a by-product, a similarity measure on
the set of extremes: the higher the probability that their latent
variables are equal, the more similar two extreme observations $X$ and
$X'$ are considered. The similarity matrix thus obtained naturally
defines a \textit{weighted graph}, whose vertices are the
anomalies/extremes observed, paving the way for the use of powerful
graph-mining techniques for community detection and visualization, see
\textit{e.g.} \cite{SCHAEFFER200727}, \cite{WICS:WICS1343} and the
references therein. Beyond its detailed description, the methodology
proposed is applied to a real fleet monitoring dataset in the
aeronautics domain and shown to provide useful tools for analyzing and
interpreting abnormal data.

The paper is structured as follows. Basic concepts of MEV theory are briefly recalled in Section \ref{sec:background}, in particular the concept of angular measure, together with the technique proposed in \cite{GoixAISTATS16,goix2017sparse} for estimating the (hopefully sparse) % angular measure
support of the latter, which determines the dependence structure of extremes arising from a  heavy-tailed distribution. Section \ref{sec:model} % introduces the concept of angular measure and
details the mixture model we propose to describe the distribution of extreme data, based on the output of the support estimation procedure,  together with the EM algorithm variant we introduce in order to estimate its parameters. It is next explained in Section \ref{sec:viz} how to exploit the results of this inference method to define a similarity matrix of the extremal data, reflecting a weighted graph structure of the observed anomalies, and apply dedicated community detection and visualization techniques so as to extract meaningful information from the set of extreme observations. The relevance of the approach we promote is finally illustrated by numerical experiments, on synthetic and real data in Section~\ref{sec:experiment}. An implementation of the proposed method and the code for the experiments carried out in this paper are available online\footnote{ \url{https://github.com/mchiapino/mevt_anomaly}}. 
Technical details are deferred to the Appendix section. 

%%% Local Variables:
%%% mode: latex
%%% ispell-local-dictionary: "american"
%%% TeX-master: "Anomaly_VisualDisplay_Long.tex"
%%% End:

\section{Background and Preliminaries}\label{sec:background}
We start with recalling key notions of MEVT,  the concept of angular measure in particular,
as well as the inference method investigated in \cite{GoixAISTATS16,goix2017sparse} to estimate its support.   Here and throughout, % the power set of any ensemble $E$ is denoted by $\mathcal{P}(E)$,
the Dirac mass at any point $x$ is denoted by $\delta_x$, the indicator function of any event $A$ by $\1\{A\}$, the cardinality of any finite set $E$ by $\vert E\vert$. Capital letters generally refer to random quantities whereas lower case ones denote deterministic  values. Finally, boldface letters denote vectors as opposed to Roman letters denoting real numbers.
\subsection{Heavy-Tail Phenomena - Multivariate Regular Variation}
\label{sec:MEVT}
Extreme Value Theory (\textsc{EVT}) describes phenomena that are not governed by an 'averaging effect' but can be instead significantly impacted by very large values.  By focusing on large quantiles rather than central statistics such as the median or the sample mean, \textsc{EVT} provides models for the unusual rather than the usual and permits to assess the probability of occurence of rare (extreme) events. Application domains are numerous and diverse, including any field related to risk management as finance, insurance, environmental sciences or aeronautics.
Risk monitoring is a typical use case of \textsc{EVT}. The reader is referred to \cite{Coles2001} and the references therein for an introduction to \textsc{EVT} and its applications. 
In the univariate setting, typical quantities of interest are high quantiles of a random variable $X$, \ie\ $1-p$ quantiles for $p\to 0$. When $p$ is of the same order of magnitude as $1/N$ or smaller, empirical estimates become meaningless. Another issue is the estimation of the probability of an excess over a high threshold $u$, $p_u = \PP(X>u)$ when few (or none) observations are available above $u$. In such contexts, \textsc{EVT} essentially consists in using a parametric model (the generalized Pareto distributions) for the tail distribution, which is theoretically justified asymptotically, \ie\ when $p\to 0$ or $u\to \infty$.
Here and throughout we place ourselves in the context where the variable of interest is regularly varying; see \cite{Resnick1987,resnick2007heavy} for a general introduction to regular variation and its applications to data analysis. In the univariate case 
the required assumption 
is the existence of a 
sequence  $a_n>0$ such that   $a_n \to \infty$ and a function $h(x)$ % and a non-degenerate cumulative distribution function (\cdf) $G$
such that $n \PP[X/a_n> x] \nto h(x)$, $x>0$. 
% \[ n ~\mathbb{P}\left( (X - b_n)/a_n  ~\ge~ x \right) \rightarrow -\log G(x)\, \text{ as }n\rightarrow \infty\,,\]
%for all $x$ in the continuity set of $G$.
Notice that this assumption is satisfied by most textbook heavy tailed  distributions,
 \eg\ %normal, exponential,
 Cauchy, Student.
 %beta, gamma distributions. The reader is referred to \cite{Coles2001} and the references therein for an introduction to \textsc{EVT} and its applications.
In such a case $h$ is necessarily of the form $h(x) = C x^{-\alpha}$ for some $C, \alpha >0$, where  $\alpha$ is called the \emph{tail index} of $X$ and $a_n$ may be chosen as $a_n = n^{1/\alpha}$. 
 In the multivariate setting,
% \textsc{EVT} is
% The focus is now on the tail
% behaviour of
 consider a $d$-dimensional random vector $\mb X = (X_1,\ldots, X_d)$,
 the goal is to infer quantities such as
 $\PP[X_1>x_1, \ldots, X_d>x_d]$ for large $x_1,\ldots,x_d$.  A
 natural first step is to standardize each marginal distribution so
 that the $X_j$'s are all regularly varying with tail index $\alpha = 1$ and scaling constant $C=1$.
 One convenient choice is to use the probability integral transform.
 For $\mb x = (x_1,\ldots,x_d)$, let $F_j(x_j) = \PP[X_j\le
 x_j]$. Assuming that $F_j$ is continuous, the transformed variable
 $V_j = (1- F_j(X_j))^{-1}$ follows a Pareto distribution,
 $ \PP[V_j> v] = v^{-1}$, $v\ge 1$. In practice $F_j$ is unknown but its empirical version $\hat F_j$ may be used instead. Another option when each $X_j$ is
 regularly varying with tail index $\alpha_j$ is to estimate   $(\alpha_j, C_j)$ using \emph{e.g.} a
 generalized Pareto model above large thresholds, see \cite{Coles2001}
 or \cite{BGTS04} and the references therein. Then
 $V_j = X^{\alpha_j}/ C_j$ is standard regularly varying, meaning that 
%\begin{equation}
 % \label{eq:standard1D_RV}
 $n \PP[V_j /n > x  ] \rightarrow x^{-1}, x>0$ as $n\rightarrow \infty$. 
%\end{equation}
% Consider the the Pareto-tranformed variable
% $\mb V = (V_1, \ldots, V_d)$.
The multivariate extension of the latter assumption is that the standardized vector $\mb V = (V_1,\ldots, V_d)$ is regularly varying with tail index equal to $1$, \ie\ there exists % a sequence $a_n$ as above and
a limit Radon measure $\mu$ on $\rset_+^d\setminus\{0\}$ such that
\begin{equation}
  \label{eq:standardRV}
n  \PP[n^{-1} \mb V \in A] \to \mu(A)  
\end{equation}
for all $A$ in the continuity set of $\mu$ such that $0\notin\partial A$.   
% {\bf Exponent  measure.}
% To understand the nature of the tail distribution assumption, the following result (see \emph{e.g.}\cite{Resnick1987,resnick2007heavy}) is key: 
% there exists a measure $\mu$ on $E = \rset_+^d\setminus\{0\}$  which is finite on any set $A$ such that $\bzero$ does not belong to the closure of $A$, 
%  such that 
% $-\log G(\mb v) = \mu [\bzero, \mb v]^c$.
The measure $\mu$ is called the \emph{exponent measure}. In the standard setting characterized by~%\eqref{eq:standard1D_RV} and~
\eqref{eq:standardRV}, it is homogeneous of order $-1$, that is $\mu(tA) = t^{-1} \mu(A)$, where $tA= \{t\mb v, \mb v \in A\}, A\subset \rset_d^+$ and $\mu\{\mb x \in\rset_+^d: x_j \ge 1\}=1$.   % Another consequence is that for all $A\subset\rset_+^d$ such that $0\notin\partial A$, $
% t\mathbb{P}(\mb V\in t A)\rightarrow  \mu(A)$ as $t\rightarrow \infty$.
Assumption~\eqref{eq:standardRV} applies immediately to the problem of estimating the probability of reaching a set $t A$ which is far form $\bzero$ (\ie\ $t$ is large): one may write
$\PP(\mb V \in t A)  \approx \frac{1}{t}\mu(A)$, so that estimates of $\mu$ automatically provide estimates for such quantities.
In  a word, $\mu$ % and $\Phi$ are in one-to-one correspondence and any one of them
may be used to characterize the distributional tail of $\mb V$. 
% $
For modeling purposes, the homogeneity property
$\mu(t\cdot ) =t^{-1}\mu(\cdot)$ suggests a preliminary decomposition
of $\mu$ within a (pseudo)-polar coordinates system, as detailed next.
\subsection{Angular Measure - Dependence in the Extremes}
% \paragraph{Angular measure}
Consider the sum-norm $\sumnorm{\mb{v}}:=v_1+\ldots+v_d$ and
$\mathcal{S}_d:=\{\mb{w}\in \rset_+^d : \sumnorm{\mb{w}}=1\}$ the
$d$-dimensional simplex.  Introduce the polar transformation
$T: \mb v\mapsto T (\mb v) = (r,\mb{w})$ defined on
$\rset_+^d\setminus \{\mb{0}\}$, where
% \begin{align*}
%   T(\mb{v})=(r,\mb{w}),
    %   \end{align*}
$r=\|\mb{v}\|$ is the radial component and
$\mb{w}={r}^{-1}{\mb{v}}$ is the angular one.  Now define the
\emph{angular measure} $\Phi$ on
$\mathcal{S}_d$ (see \emph{e.g.}  \cite{resnick2007heavy} or
\cite{BGTS04} and the references therein): $
\Phi(A):=\mu\left\{\mb{v}:\sumnorm{\mb{v}}>1,\sumnorm{\mb{v}}^{-1}\mb{v}\in
  A\}\right\}$, with $A\subset
\mathcal{S}_d$. Notice that
$\Phi(\mathcal{S}_d)<\infty$ and, by homogeneity,

\begin{align}
  \label{eq:substitution}
  \mu\circ T^{-1}(\mathrm dr, \mathrm d\mb{w})=r^{-2}\mathrm dr\Phi(\mathrm d\mb{w}).
\end{align}
In other words the exponent measure $\mu$ factorizes into a tensor product of a radial
component and an angular component. Setting $R=\|\mb{V}\|$ and
$\mb{W}= R^{-1} \mb{V}$, a consequence is that
% then a re-writing of (\ref{eq:regular-variation}) is

\begin{align}  
  \PP[\mb{W}\in A , R>tr \given R>t ] &\xrightarrow[t\to\infty]{} r^{-1} \Phi(\simplex_d)^{-1} 
                                        \Phi(A) \label{eq:ang-measure-lim}% \frac{\Phi(A)}{\Phi(\mathcal{S}_d)}. \\
                                        %% \PP[R> tr | R>t ] &\xrightarrow[t\to\infty]{} r^{-1}
\end{align}
for all measurable set $A\subset\mathcal{S}_d$ such that $\Phi(\partial A)=0$ and $r>1$. Hence,
given that the radius $R$ is large,  $R$ and the angle
$\mb W$ are approximately independent, the distribution of $\mb W$ is
approximately the angular measure -- up to a normalizing constant
$\Phi(\simplex_d)$ -- and $R$ follows approximately a Pareto
distribution. As it describes the distribution of the directions formed by the largest observations, the angular measure exhaustively accounts for the dependence structure in the extremes.
Our choice of a standard regular variation framework~\eqref{eq:standardRV} % The transformation to unit Pareto margins
and that of  the sum-norm yield the following  moment constraint on $\Phi$: 

\begin{align}
  \label{eq:mom-constraint}
  \int_{\mathcal{S}_d}w_i\, \Phi(\mathrm d\mb{w}) = 1, \text{ for } i=1,\ldots,d.
\end{align}
In addition, the normalizing constant is explicit:

\begin{align}
  % \begin{split}
  \label{eq:phi-mass}
  \Phi({\mathcal{S}_d}) =\int_{\mathcal{S}_d}\Phi(\mathrm d\mb{w})
  =\int_{\mathcal{S}_d}(w_1+\ldots+w_d)\Phi(\mathrm d\mb{w})
  =d.
  % \end{split}
\end{align}
\begin{remark}
  The choice of the sum-norm here % for defining the angular measure
  is somewhat arbitrary. Any other norm on $\rset^d$ for the
  pseudo-polar transformation is equally possible, leading to
  alternative moment constraints and normalizing constants. The
  advantage of the sum-norm is that it allows convenient probabilistic
  modeling of the angular component $\mb w$ on the unit simplex.
\end{remark}

\subsection{Support Estimation - the DAMEX Algorithm}\label{sec:support-estimation}
We now expose  the connection between $\Phi$'s (or equivalently, $\mu$'s) support and the subsets of components which may simultaneously take very large values, while the others remain small.
\medskip

\noindent {\bf Sparse support.}
Fix  $\alpha \subset \dd$ and consider the associated  truncated cone

\begin{equation}
\label{cone}
\mathcal{C}_\alpha = \big\{\mb v \ge 0 : ~\|\mb v\|_\infty \ge 1,~ v_i > 0 \text{ for } i \in \alpha, \; v_i = 0 ~\text{ for } i \notin \alpha \big\}.
\end{equation}
The family
$\{\mathcal{C}_\alpha, \alpha\subset\{1,\ldots ,d\},
\alpha\neq\emptyset\}$ first introduced in~\cite{GoixAISTATS16} defines a partition of
$\mathbb{R}_+^{d}\setminus [0,1]^d$ which is of particular interest
for our purpose: notice first that, by homogeneity of $\mu$,
the following equivalence holds: %We have the equivalence for all $\alpha\in\mathbb{M}_0$,
%\begin{align}
 % \label{eq:sparsity-mu-phi}
  $\Phi(\mathcal{S}_\alpha)>0 \Leftrightarrow \mu(\mathcal{C}_\alpha)>0$, 
%\end{align}
% Second, 
% $\mu(\cone_\alpha)>0$ if and only if $\Phi(\mathcal{S}_\alpha)>0$,
where $  \simplex_\alpha = \left\{\mb v \in\rset_+^d: \; \sumnorm{v} = 1,\; v_i >0 \text{ for } i\in\alpha,
    v_i =0 \text{ for } i\notin\alpha \right\}$,   $ \varnothing\neq \alpha\subset \dd$. Observe next that  $\mu(\cone_\alpha)>0 $ means that the
limiting rescaled probability that `all features in $\alpha$ are simultaneously large,
while the others are small' is non zero. Precisely,  consider  the
$\epsilon$-thickened rectangle

\begin{equation*}
% \label{eq:epsilonRect}
 \mathcal{R}_\alpha^\epsilon~=~\big\{\mb v \ge
 0,~\|\mb v\|_\infty \ge 1,~ v_i > \epsilon  ~\text{ for } i
 \in \alpha,\; v_i \le \epsilon ~\text{ for } i \notin \alpha
 \big\}, 
\end{equation*}
which corresponds to the event that all features in $\alpha$ are large, while the other are small. 
The $\mathcal{R}_\alpha^\epsilon$'s define again a
partition of $\mathbb{R}_+^{d}\setminus [0,1]^d$ for each fixed $\epsilon\ge 0$. In addition, we have that  $\cone_\alpha=\cap_{\epsilon>0,\epsilon\in\qset} \rect_{\alpha}^\epsilon$, so that 
 by upper continuity of $\mu$,
\[\mu(\cone_\alpha)= \lim_{\epsilon\to 0} \mu(\rect_\alpha^\epsilon) \]
% \lim_{t\to \infty }t \PP ( ) $
    with 
\[\mu(\rect_\alpha^\epsilon)
    = \lim_{t\to \infty }t \PP ( \ninf{\mb V} >t,\;
    \forall j \in \alpha:V_j > t\epsilon, \;
    \;\forall j \notin \alpha:V_j < t\epsilon). \] 
    In the sequel, set
$  \mu_\alpha= \mu(\cone_\alpha)$,
      $\mathbb{M}=\big\{\alpha\subset\dd, \alpha\neq\varnothing,
      \mu_\alpha>0\big\}$.
      Although every $\mu_\alpha$ may be positive in theory, a reasonable assumption in many practical high dimensional situations is that $\mu_\alpha = 0$ for the
vast majority 
of the $2^d -1$ cones $\mathcal{C}_\alpha$. In other words, not all combinations of coordinates of $\mb V$ can be large together, so that the support of $\mu$ (and that of $\Phi$) is sparse.
\medskip

\noindent {\bf Support estimation.}
The task of estimating the support of $\mu$ (or
$\Phi$) has recently received increasing attention in the statistics
and machine learning literature. \cite{chautru2015dimension} first
proposed a non parametric clustering approach involving principal
nested spheres, which provides great flexibility at the price of
computational cost.  In contrast, \cite{GoixAISTATS16}'s methods rely
on the above mentioned partition of the unit sphere into
$2^d-1$ sub-simplices
$\Omega_\alpha$ and led to the so-called DAMEX algorithm which
computational complexity $O(dn\log
n)$ scales well with higher dimensions.  Their algorithm produces the
list of $\alpha$'s such that the empirical counterpart of
$\mu_\alpha$ (denoted $\hat
\mu_\alpha$ in the sequel) is non zero. Defining a threshold
$m_{\min}>0$ below which $\hat
\mu_\alpha$ is deemed as negligible, one thus obtains a list of
subsets $\widehat{\mathbb{M}} = \{\alpha\subset\dd: \hat
\mu_\alpha>\mu_{\min}\}$.  A uniform boud on the error $|\hat
\mu_\alpha -
\mu_\alpha|$ is derived in~\cite{goix2017sparse} which scales roughly
as $k^{-1/2}$, where
$k$ is the order of magnitude of the number of largest observations
used to learn $\mathbb{M}$ and the
$\mu_\alpha$'s.  In \cite{simpson2018determining} the original DAMEX
framework is refined in order to also model extremes in the directions
$\Omega_\alpha$ where the angular measure does not concentrate. A
third algorithm named CLEF has been proposed
by~\cite{chiapino2016feature} which allows to cluster together
different sub-simplices
$\Omega_\alpha$'s which are close in terms of symmetric difference of
the subsets
$\alpha$'s. This is particularly useful in situations where the
empirical angular mass is scattered onto a large number of
sub-simplices, so that DAMEX fails to exhibit a list $\hat{ \mathbb{M
  }}$ of reasonable size.  Asymptotic guarantees for the latter
approach and variants, leading to statistical tests with controllable
asymptotic type I error are derived in~\cite{chiapino2018identifying}.

In the present paper, support estimation is only a preliminary step before mixture modeling. We decided to use DAMEX in view of its computational simplicity and the statistical guarantees it offers, considering the fact that its output was very similar to CLEF's on the aeronautics dataset considered in our usecase (see Section~\ref{sec:flights_visu}). Using the above mentioned alternatives is certainly possible but for the sake of brevity we shall only present the results obtained using DAMEX as a preprocessing step. We now briefly describe how DAMEX works.  
 \medskip

% Earlier works (\cite{GoixAISTATS16}) have proposed an algorithm named
% DAMEX which 

\noindent {\bf The DAMEX algorithm. } 
Given a dataset
$(\mb X_i)_{i\le n}$ of independent data distributed as
$\mb X$, DAMEX  proceeds as follows. First, replace the unknown
marginal distributions $F_j$ with their empirical counterpart
$\hat F_j(x) = \frac{1}{n}\sum \1\{X_{i,j} < x\}$ and define next
$\hat V_{i,j} = (1 - \hat F_j(X_{i,j}) )^{-1}$ and
$\mb{\hat V}_i= (\hat V_{i,1},\ldots, \hat V_{i,d})$. Then choose some $k \ll n$ large
enough (typically $k = O(\sqrt{n})$) and define $\hat \mu_\alpha$ as the empirical 
counterpart of $\mu(R_\alpha^\epsilon)$ with $t$ replaced by $n/k$, that is
$\hat \mu_\alpha = (1/k) \sum_{i=1}^n \1\{\hat{\mb V_i} \in \frac{n}{k}\rect_\alpha^\epsilon \}$.
Notice that the above description is a variant of the original algorithm in~\cite{GoixAISTATS16} which uses thickened cones $\cone_\alpha^\epsilon$ instead of  $\rect_\alpha^\epsilon$. However finite sample guarantees in \cite{goix2017sparse} are obtained using the latter rather than the original $\cone_\alpha^\epsilon$'s, that is why  using the $\rect_\alpha^\epsilon$'s is preferred.

% \todo{Elargir la revue de la littératue exixtante à CLEF (versions courtes et longues)  et au preprint de Emma Simpson}

%%% Local Variables:
%%% mode: latex
%%% ispell-local-dictionary: "american"
%%% TeX-master: "Anomaly_VisualDisplay_Long.tex"
%%% End:

\section{A Mixture Model For Multivariate Extreme Values}\label{sec:model}
The purpose of this section is to develop a novel mixture model for the angular distribution $\Phi$ of the  largest instances of the dataset, indexed by
$\alpha \in \mathbb{M}$, where $\mathbb{M}$ is $\Phi$'s support.   Each component
$\alpha\in\mathbb{M}$ of the
mixture % is indexed by some $\alpha\subset\{1,\ldots,d\}$,
generates instances $\mb V$ such that $V_j$ is likely to be large for
$j\in\alpha$ and the latent variables of the model take their values in $\mathbb{M}$. In practice, we adopt a \textit{plug-in} approach and identify $\mathbb{M}$ with $\widehat{\mathbb{M}}$, the output of
DAMEX.  As the distribution of extremes may be  entirely characterized by the distribution of their angular component $\mb W \in \simplex_d$ (see the polar decompositions~\eqref{eq:substitution} and~\eqref{eq:ang-measure-lim}), a natural model choice is that of   Dirichlet mixtures.
%%\todo{ Citer Boldi-davison, Sabourin-naveau}
%We next show how to infer the parameters of the mixture model proposed by means of a variant of the EM algorithm.
We next show how to design a 'noisy' version of the model for subasymptotic observations and how to infer it by means of an EM procedure based on a truncated version of the original dataset, surmounting difficulties related to the geometry of $\Phi$'s support.
\subsection{Angular Mixture Model for the Directions along which Anomalies Occur}\label{sec:mixtureModel}%% Model for Multivariate Extreme Values}
% The partition of $\rset_+^d$ into cones $\cone_\alpha$ introduced in Section~\ref{sec:support-estimation} 
% induces a
Recall from Section~\ref{sec:support-estimation} that $\simplex_d$ is naturally partitioned  into $2^d - 1$ sub-simplices $\simplex_\alpha$. 
% $  \simplex_\alpha = \left\{\mb v \in\rset_+^d: \; \sumnorm{v} = 1, v_i >0 \text{ for } i\in\alpha,
%     v_i =0 \text{ for } i\notin\alpha,\right\}$,   $ \varnothing\neq \alpha\subset \dd$. 
  % Recall % from Section~\ref{sec:support-estimation}
% that o
Our key assumption is that the support of $\mu$ (or $\Phi$) is \textit{sparse} in the sense that $|\mathbb{M}|\ll 2^d$, where $\mathbb{M} = \{\alpha:\mu(\cone_\alpha)>0 \} = \{\alpha: \Phi(\simplex_\alpha)>0\}$. 
Let  $K$ denote the number of subsets $\alpha\in\mathbb{M}$ of cardinality at least $2$ and let $d_1\in\{0,\ldots,d\}$ be the number of  singletons $\{j\}\in\mathbb{M}$. Without loss of generality we assume that these singletons correspond to the first $d_1$ coordinates, so that $\mathbb{M} = \{\alpha_1,\ldots, \alpha_K, \{1\},\ldots, \{d_1\}\}$. 
For simplicity, we also suppose that the sets $\alpha\in\mathbb{M}$ are not nested, % , \ie\
% there does not exist two subsets $\alpha,\beta\in\mathbb{M}$ such that
                                                               %                                                                $\alpha \subset\beta$. Notice that this
an hypothesis which can be relaxed at
the price of additional notational complexity.
In view of \eqref{eq:phi-mass}, the angular measure then admits the decomposition

% \begin{align}\label{eq:mixture-Phi}
%   \Phi(\point) =  d \sum_{\alpha\in\mathbb{M}} \pi_\alpha \Phi_\alpha(\point),
% \end{align}
% where $\Phi_\alpha$ is a probability measure on $\simplex_\alpha$ and
% $\sum_{\alpha\in\mathbb{M}} \pi_\alpha = 1$.  We make the simplifying
% assumption that the sets $\alpha\in\mathbb{M}$ are not nested, % , \ie\
% % there does not exist two subsets $\alpha,\beta\in\mathbb{M}$ such that
%                                                                %                                                                $\alpha \subset\beta$. Notice that this
% an assumption which could be omitted at
% the price of additional notational complexity.
% Introduce the set of coordinates % $j\in\{1,\ldots, d\}$
% which are singletons in $\mathbb{M}$, % subset of $\matbb{M}$ made of singletons
% $\mathbb{M}_1 = \{ j \in \dd: \{j\}\in\mathbb{M} \}$, as opposed to
% $\mathbb{M}_2= \{\alpha\in\mathbb{M}: |\alpha|\ge 2\}$.  Up to
% relabeling we may assume that $\mathbb{M}_1 = \{1,\ldots, d_1\}$ for some
% $1\le d_1\le d\}$ or that $\mathbb{M}_1=\varnothing$, in which case
% $d_1=0$. Then
% $\bigcup_{\alpha\in\mathbb{M}_2} \alpha = \{d_1+1,\ldots, d\}$.  For
% convenience let us write $\mathbb{M}_2=\{\alpha_1,\ldots,\alpha_K\}$
% with $K=|\mathbb{M}_2|$ and let us relabel the weights as
% $\pi_k = \pi_{\alpha_k}$ for $k\le K$, $ \pi_{K+j} = \pi_{\{j\}}$ for
% $j\le d_1$. Equipped with these notations,~\eqref{eq:mixture-Phi}
% becomes
% \begin{equation}
%   \label{eq:mixture-Phi2}

\[d^{-1} \Phi(\point) =  \sum_{k=1}^K \pi_k \Phi_{\alpha_k}(\point) + \sum_{j\le d_1 } \pi_{K+j} \delta_{\mb e_j}(\point),\]
%\end{equation}
where $\Phi_{\alpha_k}$ is a probability measure on $\simplex_{\alpha_k}$, the weights  $\pi_k$  satisfy $\sum_{k\le K+j} \pi_k = 1$ and $\mb{e}_j=(0,\ldots,1,\ldots,0)$ is the $j^{th}$ canonical basis vector of $\rset^d$.
% Weights are assigned to each component of the model, write $\mb{p}=(\pi_1,\ldots,\pi_K)$
% and $(e_j)_{j\in\mathbb{E}}$ all positive such that $\sum_{k=1}^K\pi_k+\sum_{j\in\mathbb{E}}e_j=1$.
The singletons weights derive immediately from  the moment constraint~\eqref{eq:mom-constraint}: for $i\le d_1$, 

\[d^{-1}
  = % \underbrace{
    \sum_{k=1}^K\int_{\mathcal{S}_{\alpha_k}} w_i  \pi_k \Phi_{\alpha_k}(\ud \mb w) %}_{=0}
  + \sum_{j\le d_1}\int_{\mathcal{S}_{\{j\}}} w_i  \,  \pi_{K+j} \delta_{\mb e_j}(\ud \mb w)
  =      \pi_{K+i}.\]
We obtain

\begin{equation}
  \label{eq:mixture-Phi2}
  \Phi(\point) =  d \sum_{k=1}^K \pi_k \Phi_{\alpha_k}(\point) + \sum_{j\le d_1} \delta_{\mb e_j}(\point), 
  \end{equation}
where  the vector $\mb{\pi} \in [0,1]^{K+d_1}$ must satisfy
\begin{equation}
  \label{eq:constraintP}
  \sum_{k=1}^K\pi_k=1- d_1/d.   
\end{equation}
                                         %                                                                This is not properly a 'mixture' of densities as they are not defined on the same subspaces.
                                                               %                                                               
% $\tilde{\mb{v}}$ has been generated by the underlying cluster $\alpha_k$ (resp by the $j$-th feature) while the other coordinates are $0$.
%\subsection{A mixture model for the largest observations}
                                                               %                                                                {\bf Dirichlet mixture angular model.}
Equation~\eqref{eq:mixture-Phi2} determines the structure of the
angular distribution of the largest observations. For likelihood-based inference, a parametric model for each component $\Phi_{\alpha_k}$ of the
angular measure must be specified.  One natural model for
probability distributions on a simplex is the Dirichlet family, which
provides a widely used prior in Bayesian statistics for data clustering purposes in particular.
We recall that the
Dirichlet distribution on a simplex $\simplex_\alpha$ admits a density
$\diri_\alpha$ with respect to the $(|\alpha|-1)$-dimensional Lebesgue measure
which is denoted by $\ud \mb w$ for simplicity. It can be parameterized by
a mean vector
                                                               %                                                                
                                                               %                                                                
                                                               %                                                                Therefore for  $\alpha\in\mathbb{M}$ 
                                                               %                                                                we set a Dirichlet distribution on $\mathcal{S}_\alpha$. It can be parametrized by a mean vector
$\mb{m}_\alpha \in \simplex_\alpha$
and a concentration parameter $\nu_\alpha>0$, so that for $ \mb{w}\in\mathcal{S}_\alpha$,

\begin{align*}
%  \label{eq:dir-density}
  \diri_\alpha(\mb{w} | \mb{m}_\alpha,\nu_\alpha)=\frac{\Gamma(\nu_\alpha)}{\prod_{i\in\alpha}\Gamma(\nu_\alpha m_{\alpha,i})}\prod_{i\in\alpha}w_i^{\nu_\alpha m_{\alpha,i} - 1}. 
\end{align*}
Refer to
\emph{e.g.} \cite{muller2004nonparametric} for an account of
Dirichlet processes and mixtures of Dirichlet Processes applied to Bayesian nonparametrics. We emphasize that our context
is quite different: a Dirichlet Mixture is used here as a model for the angular component of the largest observations, not as
a prior on parameters. This modeling strategy for extreme values was first proposed in~\cite{boldi2007mixture} and revisited in~\cite{sabourin2014bayesian} to handle the  moment constraint \eqref{eq:mom-constraint}  via a model re-parametrization. In both cases, the focus was on moderate dimensions. In particular, both cited references worked under the assumption that the angular measure concentrates on the central simplex $\Omega_{\{1,\ldots, d\}}$ only. In this low dimensional  context, the main purpose of the cited authors was to derive the posterior predictive angular distribution in a Bayesian framework, using a variable number of mixture components concentrating on $\Omega_{\dd}$. Since the set of Dirichlet mixture distributions with an arbitrary number of components is dense among all probability densities on the simplex, this model permits in theory to approach any angular measure for extremes. 
The scope of the present paper is different. Indeed we are concerned with high dimensional data (say $d \simeq 100)$ and consequently we do not attempt to model the finest details  of the angular measure. Instead we intend to design a model accounting only for  information which is relevant for clustering. Since an intuitive summary of an extreme event in a high dimensional context is the subset $\alpha$ of features   it involves,  we assign one mixture component per sub-simplex $\Omega_\alpha$ such that $\alpha \in\mathbb{M}$. Thus we model each % consider of mixture of Dirichlet distributions $ \Phi_\alpha$ 
% is paper we model
$\Phi_\alpha$ by a single Dirichlet distribution with
unknown parameters $m_\alpha, \nu_\alpha$.  Using the standard fact that for such a distribution, 
$\int_{\mathbb{S}_\alpha} \mb w \diri_\alpha(\mb w |m_\alpha,
\nu_\alpha) \ud \mb w = \mb m_\alpha$,
% For all $j\in\{1,\ldots,d\}\setminus\mathbb{E}$
the moment constraint~(\ref{eq:mom-constraint}) becomes:

\begin{align}
  % \begin{split}
  \label{eq:mom-constraint-dir}
  \frac{1}{d} % &= \int_{\mathcal{S}_{supp}}w_j\, \frac{\Phi(\mathrm d\mb{w})}{\Phi(\mathcal{S}_{supp})}\\
  % &= \sum_{k=1}^K\int_{\mathcal{S}_{\alpha_k}}w_j\, \frac{\Phi(\mathrm d\mb{w})}{\Phi(\mathcal{S}_{supp})}\\
  % &=\sum_{k=1}^K\pi_k\int_{\Delta_{\alpha_k}}I_k(\mb{u})_j\, Dir_{\alpha_k}(I_k(\mb{u}) | \mb{m}_k,\nu_k)\mathrm du_1\ldots\mathrm du_{|\alpha_k|-1}\\
                &=\sum_{k=1}^K\pi_k\mb{m}_{k,j},  \quad j \in \{d_1+1,\ldots,d\},
                % \end{split}
\end{align}
where $\mb m_k = \mb{ m}_{\alpha_k}$ for $k\le K$. 
                  %                   with $\Delta_{\alpha_k}=\{\mb{u}\in (0,\infty)^{|\alpha_k|-1}:u_1+\ldots+u_{|\alpha_k|-1}=1\}$ and
                  %%                   $I_k$ such that $I_k(u)\in\mathcal{S}_{\alpha_k}$.%$
                  %                   This model is summarized in the Supplementary Material.

\subsection{ A Statistical Model for Large  but Sub-asymptotic Observations.}
\begin{figure}[hbtp]
  \centering
  \includegraphics[width=0.5\linewidth]{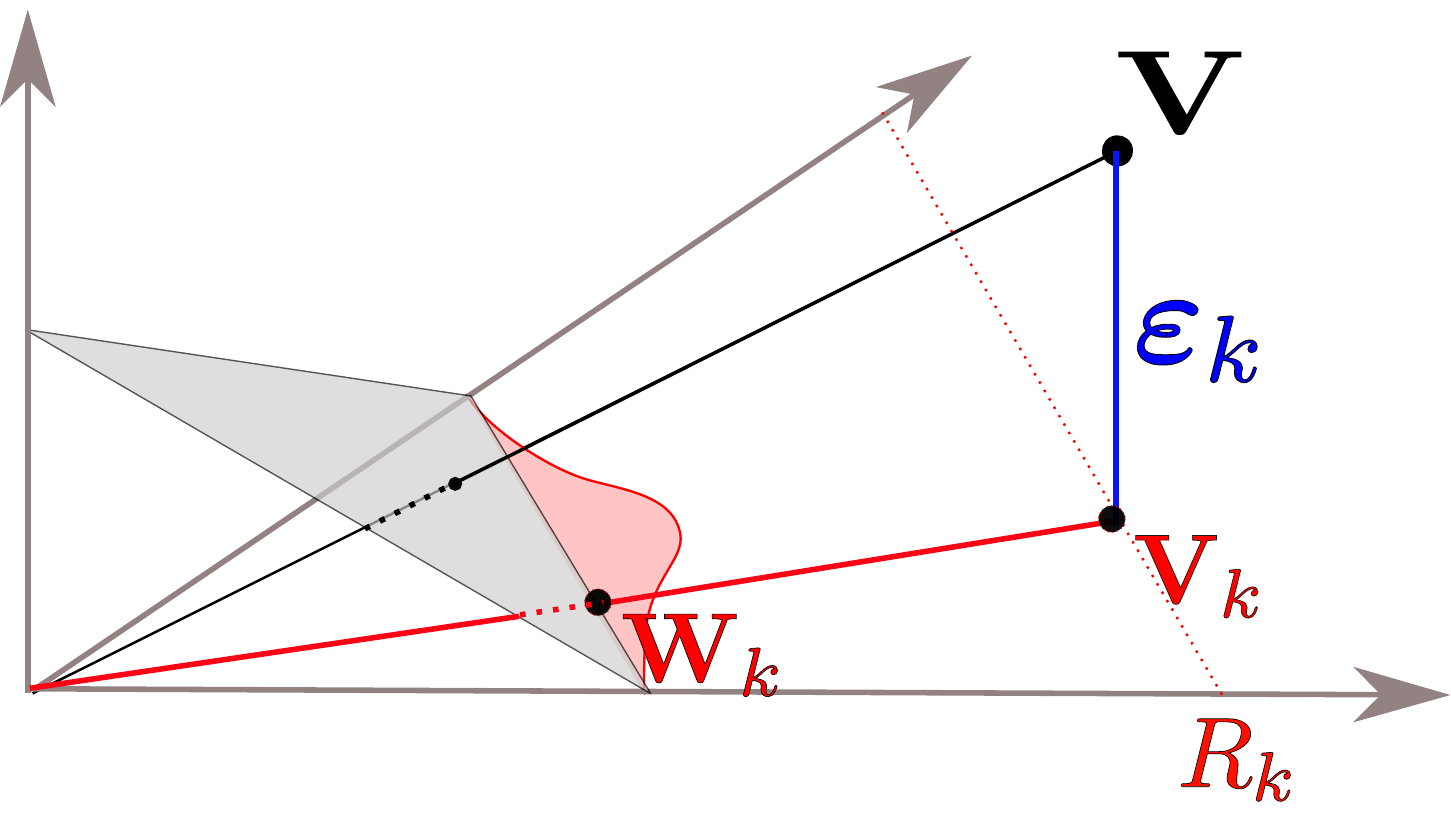}%%figures/noisyModel_CSDA2018}
  \caption{Trivariate illustration of the sub-asymptotic model~\ref{model:noisy}: \newline the observed point ${\mb V}$ has been generated by component $\alpha_k = \{1,2\}$. The grey triangle is the unit simplex, the shaded red area stands for the Dirichlet density $\varphi_k$.
}
  \label{fig:noisyModel}
  \end{figure}

Recall from~\eqref{eq:ang-measure-lim} that $\Phi$ is the
\emph{limiting} distribution of $\mb V$ for large $R$'s. In practice, we dispose of no realization of this limit probability measure and
the observed angles corresponding to radii $R>r_0$  % for $r_0$ a large
                  %                   threshold fixed by the user
follow a sub-asymptotic version of $\Phi$.
In particular, if the margins   $V_j$ have a continuous distribution, % , the marginal variables $V_j$ have continuous Pareto
% distributions, so that $\PP(V_j >1 )=1,
% \,j\in\dd$. % Also by construction, their empirical versions satisfy $\hat V_i^j\ge 1$.
%As a consequence with probability $1$, all the 
we have $\PP(V_j \neq 0 )=1$ so that  all the $\mb{V}_i = (V_{i,1}, \ldots, V_{i,d} )$, $1\le i\le n$, lie in the central cone $\cone_{\dd}$ 
% and all the angles $\mb W_i = (W_{i,1}, \ldots W_{i,d})$ lie in $\simplex_{\dd}$, the interior of
% $\simplex_d$
(this is also true using the empirical versions
$\hat{\mb V}_i$ defined in subsection~\ref{sec:support-estimation}).
%To account for the non-asymptotic nature of the data, we model the
In the approach we propose, the deviation of $\mb V$ from its  asymptotic support, which is
$\bigcup_{\alpha\in\mathbb{M}} \cone_\alpha$, is accounted for by  a noise $\bepsilon$
with light tailed distribution, namely an exponential distribution. That is,  we assume that $\mb V = R \, \mb{W} + \bepsilon$, see Model~\ref{model:noisy} below.
% We
% denote by $\tilde{ \mb V} = \mb V + \bepsilon$ the resulting noisy
% vector and we assume that only $\tilde{\mb V}$ is observed (not $\mb V$).
As is usual for mixture modeling purposes, we introduce a multinomial latent variable $\mb{Z}=(Z_1,\; \ldots,\; Z_{K+d_1})$  such that $\sum_kZ_k=1$ and %for $k\le K$ (\emph{resp.} $k >K$) ,
$Z_k=1$ if  $\mb W$ has been generated by the $k^{th}$ component of the angular  mixture~\eqref{eq:mixture-Phi2}. In a nutshell, the type of anomaly/extreme is encoded by the latent vector $\mb{Z}$. 
 % component $\Phi_{\alpha_k}$  (\emph{resp.} $\delta_{\mb e_{k-K}}$) and $Z_k=0$  otherwise.
 Then, for $k\le K$, 
$ \PP[Z_k= 1] = \pi_k$,  while, for $K< k \le K+d_1$,   $ \PP[Z_{k} = 1] = d^{-1} $.
The unknown parameters of the model are
  $\btheta = (  \mb\pi , \mb m ,
  \boldsymbol{ \nu})$, where $ \nu_k>0$ and $\mb\pi=(\pi_1,\ldots, \pi_K),\; \mb m=(\mb m_1,\ldots, \mb m_K)$ must satisfy the
  constraints~\eqref{eq:constraintP} and \eqref{eq:mom-constraint-dir}, 
 as well as  the exponential rates $\blambda = (\lambda_1, \; \ldots,\; \lambda_{K+d_1})$, where $\lambda_k>0$. 
Figure~\ref{fig:noisyModel} % in the Supplementary Material
illustrates Model~\ref{model:noisy} in dimension $d=3$.

\begin{center}
\fbox{\begin{minipage}[t]{\linewidth}
\medskip

\begin{model}[Sub-asymptotic mixture model]\label{model:noisy}~
  \begin{itemize}
  \item Consider a standard regularly varying random vector ${\mb V}$ satisfying \eqref{eq:standardRV} % with tail index $1$  which margins satisfy $\lim_{t\to \infty} t\PP(V_j> t) = 1$, $j=1,\ldots, d$  
  % are approximately Pareto
  (typically $V_j = (1 - \hat F_j(X_j))$ for $\hat F_j$ an estimate of the marginal distribution $F_j$ of $X_j$, see subsection~\ref{sec:MEVT}).
  \item Let 
    $ R = \sumnorm{{ \mb
        V}}$. Fix some high radial threshold $r_0$, typically  a large quantile of the observed radii.  Let $\mb Z$ be a hidden variable indicating the mixture component in~\eqref{eq:mixture-Phi2}. % .  responsible for the angle $\mb W$, and $k$ such that $Z_k=1$.
      % that is $\Phi_k$ if $Z_k =1$ for some $k\le K$ or $\delta_{k-K}$
 %   Fix $r_0>0$ a large threshold.
    % $\mb Z\in\{0,1\}^{K+d_1}$ be the hidden variable % as in
    % % Model~\ref{model:DM}
    % and let
%    Then %for $1\le k \le K+d_1$,
    Conditionally to
    $\{ R> r_0, Z_k=1\}$, %the observed vector
    $  {\mb{V} }$ decomposes as
    \begin{align}\label{eq:decompose-noise}
      {\mb{V}} \; = \; \mb{V}_k+ \bepsilon_k \; = \; 
      R_k \mb{W}_k+ \bepsilon_k, 
    \end{align}
    where  $\mb V_k \in \cone_{\alpha_k}$,  $\bepsilon_k\in\mathcal{C}_{\alpha_k}^{\perp}$,     $R_k=\|\mb{V}_k\|$,  $\mb{W}_k = R_k^{-1} \mb V_k \in\mathcal{S}_{\alpha_k}$.  The components $R_k, \mb W_k, \bepsilon_k$ are independent from each other.  
    The noise's components are \emph{i.i.d.} according to a translated  exponential distribution with  rate  $\lambda_k$, $R_k$ is Pareto distributed above $r_0$ and $\mb W_k$ is distributed as $\Phi_k$, that is 
%    : for $j \in \dd\setminus \alpha_k$, $
 %     \varepsilon_j\sim 1 + \dexp(\lambda_k)$.
%    are independent from each other and where 
        \begin{equation*}
          % \label{eq:noNoise-model}
\left\{          \begin{aligned}
    &\PP[R_k>r] = r_0r^{-1}, r>r_0  \,, \\
    & \mb W_k \sim \Phi_k\,,            \\
             &\varepsilon_j \sim 1 + \dexp(\lambda_k), j \in \dd\setminus\alpha_k\,,  
          \end{aligned}\right.
      % \quad \mb W_k \indep R_k
    \end{equation*}
 with $\Phi_k = \diri_k(\point|m_k, \nu_k)$ if $k\le K$, and  $\Phi_k = \delta_{\mb e_{k-K}}$ if $K<k\le K+d_1$.  
    % are as in Model~\ref{model:DM},
    % \ie\  $R_k$ is  Pareto distributed, $\mb W_k\sim \Phi_k$ and $R_k, \mb W_k$ are independent. \\
%\item

  \end{itemize}
  \medskip

 \end{model}
 \end{minipage}
}
 \end{center}

\medskip

\section{Statistical Inference via EM Algorithm.}
 In the mixture model setting described above with hidden variables
  $Z_i$, likelihood optimization is classically performed using an EM
  algorithm \citep{dempster1977maximum}. This method consists in
  performing in turn the so-called E-step and M-step at each iteration
  $t$. Denoting by $\btheta_t$ the value at iteration $t$ of the set of
  unknown model parameters, the posterior probabilities

  $$\gamma_{i,k}^{(t+1)} = \PP(Z_{i,k} = 1 | \mb V_{i},\btheta_t)$$ are
  computed during the E-step and  define the objective function

  \[Q(\btheta, \gamma^{(t)}) = \sum_{i}\sum_{k} \gamma_{i,k}^{(t+1)}\log
  p(\mb V_i | Z_{i,k} = 1, \btheta).\]
  The latter serves as a proxy for the
  log-likelihood and can be maximized with respect to $\btheta$ with
  standard optimization routines during the M-step, which yields
  $\btheta_{t+1} = \argmax_{\btheta} Q(\btheta, \gamma^{(t)})$. The procedure stops when the value  $Q(\btheta_t,\gamma^{(t)})$ reaches a stationary point and the latest pair $(\btheta_t, \gamma^{(t)})$ is returned.    
% Due to space limitations, the description of the implementation of the EM algorithm for Model~\ref{model:noisy} is postponed to the Supplementary.
% In particular, it is explained at length how to take into account the linear constraint on $(\bpi, \mb m)$ \emph{via} an adequate re-parametrization  
% which makes the M-step tractable. 

The likelihood for Model~\ref{model:noisy},
$ p({\mb{v}}|\btheta = (\mb{m},{\bnu},\mb{\pi},\blambda))$, for one
observation
${\mb v} \in (1,\infty)^d, \sumnorm{{\mb v}}\ge r_0$,
follows directly from the model
specification, % \todo{attention ! l'exposant de $r_k$ a chang\'e ($-2 \leftarrow -|\alpha_k| -1$)}

\begin{multline}
    \label{full-density-0}
    p({\mb{v}}|\btheta) %%\mb{m},{\bnu},\mb{\pi},\blambda)
    = r_0\sum_{k=1}^K\pi_k\, r_k^{-|\alpha_k|-1}\, \diri_k(\mb{w}_k|\mb{m}_k,\nu_k) 
    \prod_{j\in\alpha^c_k}  \dnoise( v_j|\lambda_k) \\%\lambda_k e^{-({{v}}_j-1) }\\%% \dexp_{\lambda_k}({\mb{v}}_j-1)
    + \frac{r_0}{d}\sum_{k = K+ 1}^{K+d_1}r_{k}^{-2} 
    \prod_{ j
      \in\{1,\ldots,d\}\setminus k} \dnoise( v_j|\lambda_k)% \lambda_k e^{-({{v}}_j-1) },
\end{multline}
where $\dnoise(\point\;|\lambda_k)$ denotes the marginal density of
the noise $\bepsilon_k$ given the noise parameter $\lambda_k$. As
specified in Model~\ref{model:noisy}, in this paper we set
$\dnoise(x|\lambda_k)= \lambda_ke^{-\lambda_k(x-1)}$, $x>1$ (a
translated exponential density), but any other light tailed
distribution could be used instead.  Notice that the term
$r_k^{-|\alpha_k| - 1} = r_k^{-2}r_k^{-|\alpha_k|+1}$ is the product
of the radial Pareto density and the Jacobian term for the change of
variables $T_k: \mb V_k\mapsto (R_k, \mb W_k)$.
Recall that the constraints are
\begin{equation}
  \label{eq:positiveConstraints}
  \nu_k>0 \; (1\le k\le K) \;, \qquad \lambda_k>0 \;(1\le k \le K+d_1), 
\end{equation}
and  that $\mb{\pi}=(\pi_1,\ldots,\pi_K)$
and $\mb m = (\mb m_1,\ldots,\mb{m}_K)$ satisfy~\eqref{eq:constraintP}
and~\eqref{eq:mom-constraint-dir}. The latter linear constraint on
$(\bpi, \mb m)$ implies that $\mb m$ and $\bpi$ cannot be optimized
independently, which  complicates the M-step of an EM-algorithm.  % (since
      %       $\mb m$ and $\bpi$ may jeopardize convergence of an EM-algorithm,
Thus we 
begin with a re-parametrization of the model ensuring that the moment
constraint~\eqref{eq:mom-constraint} is automatically satisfied. 
\medskip

\noindent {\bf Re-parametrization.} %% of the moment constraint.}
% The main idea behind the re-parametrization is to work with
% the parameter $\rho_{k,j} = \pi_km_{k,j}$ instead of $(\pi_k, m_{k,j})$.
%% \todo{ citer Sabourin-naveau, Boldi-Davison}
In a lower dimensional Bayesian framework, earlier works (\cite{sabourin2014bayesian}) have proposed a re-parametrization of the pair $(\bpi, \bf m)$ ensuring  that the moment constraint~\eqref{eq:mom-constraint} is automatically satisfied. This consists in a sequential definition of the mixture centers $m_k$ together with the involving partial barycenters of the remaining components $(m_{k+1}, \ldots, m_ k)$. The advantage if this construction is that the resulting parameter has a intuitive interpretation which facilitates the definition of a prior, while allowing for efficient MCMC with reversible jumps sampling~(\cite{green1995reversible})  of the posterior distribution. However, how to adapt    this re-parameterization to our context where several sub-simplices are involved remains an open question and we did not pursue this idea further. 
The re-parametrization that we propose here consists in working with the product parameter
 $\rho_{k,j} = \pi_km_{k,j}$ instead of the pair  $(\pi_k, m_{k,j})$. 
Namely, consider  a  $K\times (d-d_1)$ matrix $\brho = ({\brho_1}^\top,\ldots,{\brho_K}^\top) $ where $\rho_{k,j}>0$ for $j \in\alpha_k$ and $\rho_{k,j} = 0$ otherwise.
Then, for all $k\in\{1\ldots,K\}$, set

\begin{align}\label{eq:re-parametrize}
    \pi_k:=\sum_{j\in\alpha_k}\rho_{k,j} \text{ and }
    m_{k,j}:=\frac{\rho_{k,j}}{ \pi_k} %\sum_{l\in\alpha_k}\rho_{k,l}}
    , \forall j\in\alpha_k. 
\end{align}
Then \eqref{eq:constraintP} and \eqref{eq:mom-constraint-dir} together are equivalent to
%% Assume that $\mb \rho$ satisfies the constraints 
\begin{align}
  \label{eq:rho-constraint}
  \sum_{\{k : j \in \alpha_k\}}{\rho_{k,j}}& =\frac{1}{d}, \quad \forall j \in \{d_1+1,\ldots,d\}.
\end{align}

In the sequel we denote respectively by
$p({\mb v} |\brho, \bnu, \blambda) := p({\mb v} |\bpi, \mb m, \bnu,
\blambda)$ and
$\diri_k(\mb w | \brho_k, \nu_k):= \diri_k(\mb w | \mb m_k, \nu_k)$
the likelihood and the Dirichlet densities in the re-parameterized
model, where $(\mb m, \bpi)$ are obtained from $\brho$
\emph{via}~\eqref{eq:re-parametrize}.  By abuse of notations, let
$ \btheta$ denote in the sequel the set of parameters of the
re-parameterized version of Model~\ref{model:noisy}, that is
$\btheta=( \brho, \bnu, \blambda)$, and let $\Theta$ be the parameter
space, that is the set of $\btheta$'s such that
constraints~\eqref{eq:positiveConstraints}
and~\eqref{eq:rho-constraint} hold.

\medskip

\noindent {\bf EM algorithm.}
% Recall from Section~\ref{sec:model} the hidden variable 
% $\mb{Z}=(Z_1,\ldots,Z_{K+d_1})$ such that $Z_k=1$  when ${\mb{V}}$ has been generated by the $k^{th}$ mixture underlying component $\alpha_k$ (resp by the $j$-th feature) while the other coordinates are $0$.
% For a data set ${\mb V}_i, i\le n $ of independent realizations of ${\mb V}$  following Model~\ref{model:noisy},
We summarize below the EM algorithm in our framework.  Let $n_0\le n$ be 
the number of observations ${\mb V}_i$ such that
$\sumnorm{{\mb V}_i} > r_0$. To alleviate notations, we may
relabel the indices $i$ so that these observations are
$ {\mb V}_{1:n_0} =({\mb V}_1, \ldots,{\mb
  V}_{n_0})$.
Let $\mb Z_i = (Z_{i, 1}, \ldots, Z_{i, K+d_1}), i\le n_0$ be the hidden variables associated with ${\mb V}_{1:n_0}$. 
 Also let $p({\mb v}| \btheta, z_k=1 )$ denote the % likelihood for one observation given  $\btheta$ and  $z_k=1$.
conditional density of ${\mb{V}}$ given $( Z_k=1, \btheta)$.  % is involved in the expression of the lower bound $Q(\btheta, \gamma)$.
In view of the likelihood~\eqref{full-density-0}, it is given  
%for $k\le K$,
by  % \todo{attention ! l'exposant de $r_k$ a chang\'e ($-2 \leftarrow -|\alpha_k| -1$)}
\begin{equation} \label{eq:VgivenZ}
    p({\mb{v}}|z_k=1, \btheta) = 
\begin{cases}
r_k^{-|\alpha_k| -1}\, \diri_k(\mb{w_k}|\mb{\rho_k},\nu_k)\,\prod_{j\in\alpha_k^c}\dnoise( v_j|\lambda_k), %%_{\lambda_k}({\mb{v}}_j-1)
\;  (k\le K)\\
% \end{align} 
% \begin{align}
 {v}_k^{-2}\, \prod_{j \in\{1,\ldots,d\}\setminus k} \dnoise ( v_j|\lambda_k), %%Exp_{\lambda^{'}_j}({\mb{v}}_l-1)
 \quad(K<k\le K+d_1).  %\label{eq:VgivenZsingle} 
\end{cases}  
\end{equation}
% \begin{align}
%   p({\mb{v}}|z_k=1, \btheta)& =r_k^{-|\alpha_k| -1}\, \diri_k(\mb{w_k}|\mb{\rho_k},\nu_k)\,\prod_{j\in\alpha_k^c}\dnoise( v_j|\lambda_k) %%_{\lambda_k}({\mb{v}}_j-1)
%    \text{ for } k\le K \label{eq:VgivenZ}\\
% % \end{align} 
% % \begin{align}
%    p({\mb{v}}|z_k=1, \btheta)&={v}_k^{-2}\, \prod_{j \in\{1,\ldots,d\}\setminus k} \dnoise ( v_j|\lambda_k)%%Exp_{\lambda^{'}_j}({\mb{v}}_l-1)
%   \text{ for } K<k\le K+d_1.  \label{eq:VgivenZsingle}
% \end{align}
% \medskip

\begin{center}
\fbox{
\begin{minipage}[t]{\linewidth}
\medskip

%\begin{algorithm}[H]
{  \large{\bf EM algorithm for Model~\ref{model:noisy}\label{algo:EM}~}}
  \medskip
  
{\bf Input} Extreme standardized data  ${\mb V}_{1:n_0}$.
  \begin{itemize}
  \item {\bf Initialization}  Choose a starting value for $\btheta$ (See Remark~\ref{rem:initEM}).
  \item {\bf Repeat until convergence}:\\
  
    {\bf E-step}: compute  for $1\le i\le n_0$ and $k\le K+d_1$, 
      $
        \gamma_{i,k} = \PP[Z_{i,k}  =1 \given { \mb V}_i, \btheta] $
      according to \eqref{eq:gamma-ik}. Set $\bgamma = (\gamma_{i,k})_{i\le n_0, k\le K+d_1}$.
      
    {\bf M-step}: Solve the optimization problem
      $
      \max_{\btheta \in \Theta} Q(\btheta, \bgamma) 
      $%  \]
      where $
        Q(\btheta, \bgamma) = \sum_{i = 1}^{n_0} \sum_{k=1}^{K+d_1} \gamma_{i,k}\big(\log  \pi_k +  \log p({\mb V}_i | \btheta, z_{i,k}=1 )\big)$ is a lower bound for the likelihood and
        $ \pi_k = \PP(Z_{i,k} = 1|  \btheta)$, \ie\
        
      \begin{equation}
        \label{eq:pZk}
        \pi_k = % \PP(Z_{i,k} = 1|  \btheta) =
        \begin{cases}
          \sum_{\ell \in\alpha_k} \rho_{k,l}& \text{ for } 1\le k\le K \;, \\
          d^{-1} & \text{ for }  K < k \le K+d_1 \; , 
        \end{cases}
      \end{equation}
      % $ \pi_k = \pi_k$ for $k\le K$ and $ \pi_k = d^{-1}$ for $K< k\le K+d_1$; 
      where $p( {\mb V}_i | \btheta, z_{i,k}=1 )$ is given by~\eqref{eq:VgivenZ}.%-\eqref{eq:VgivenZsingle}.
      Denote by $\btheta^\star$ the solution, set $\btheta = \btheta^\star$.
  \end{itemize}
  
%\end{algorithm}
\medskip

 \end{minipage}
}
 \end{center}

 \begin{remark}\label{rem:initEM}
   In this work the starting values for  the concentration parametrers $\nu_k$ are set to $20$, those for the exponential rates are set to $\lambda_k = 0.01$. Finally, one may easily construct  a matrix $\brho$ satisfying the constraint~\eqref{eq:rho-constraint} starting with any matrix $\tilde\brho$ such that $\tilde \rho_{k,j} = 0$ for $j\notin\alpha_k$ and $\tilde \rho_{k,j}>0$ otherwise, and then defining $\brho$ via $\rho_{k, j} = (\sum_{l=1}^K \tilde \rho_{l,j})^{-1} \tilde \rho_{k,j}$. 
% one may easily 
%    it ny matrix $\brho\in\rset$  and weights $\bpi$ are chosen uniform (\ie\ $\pi_k = 1/K$), and the  
%    starting value for $\rho$ is chosen using empirical summaries of extreme data as follows: first, the
   %
 %  In this work,  the output of DAMEX is used for choosing the initial value for $\mb \rho$.
%   Namely, given $\widehat{\mathbb{M}}_2$ we compute the empirical means $\widehat m_{k,j}:=\frac{1}{n_0}\sum_{i=1}^{n_0}{ \mb V}_{i,j}$
%   for all $j$ in $\alpha_k$ and $k$ in $\{1,\ldots,K\}$ and we set $\widehat\pi_0=\ldots=\widehat\pi_K=\frac{1}{K}$ so that we get the corresponding $\widehat{\mb \rho}$
%   by $\widehat\rho_{k,j}=\pi_k m_{k,j}$. Although it is not likely to verify \eqref{eq:rho-constraint}, we can easily project $\widehat{\mb \rho}$ on $\Theta$ : ${\widehat\rho_{k,j}}^{init}=\frac{\widehat\rho_{k,j}}{d\sum_{h=1}^K\widehat\rho_{h,j}}$.
% \
\end{remark}
We now describe at length the E-step and the M-step of the algorithm. 
\medskip

\noindent \textbf{E-step.} %% This step consist in determining $(\gamma_{i,1},\ldots,\gamma_{i,K},\gamma_{i,e_1},\ldots,\gamma_{i,e_{\mathbb{E}}})$ for $i=1,\ldots,n$.
The $\gamma_{i,k}$'s are obtained using the Bayes formula, 
for  $1\le k\le K+d_1$, 
\begin{equation}
  \gamma_{i,k}=p(Z_{i,k}=1|{\mb{V}}_i,   \btheta)  %\mb{\rho}^{old},\mb{\nu}^{old},\mb{\lambda}^{old})\\
              =\frac{% p(Z_{i,k}=1)
                \pi_k \; p({\mb{V}}_i|z_{i,k}=1, \btheta)}%\mb{\rho}^{old},\mb{\nu}^{old},\mb{\lambda}^{old})}
              % {\sum_{\ell\in\{1,\ldots,K+d_1\}\setminus k}
               { \sum_{\substack{1\le \ell\le K+d_1\\ \ell\neq k}} \; %%^{K+d_1}
  \pi_\ell\; %p(Z_{i,l}=1)
  p({\mb{V}}_i|z _{i,\ell}=1,\btheta) } ,  \label{eq:gamma-ik}
\end{equation}
where $ \pi_{k}$ is defined in~\eqref{eq:pZk} and $p({\mb{V}_i}|Z_{i,k}=1, \btheta)$ is given by~\eqref{eq:VgivenZ}. % and~\eqref{eq:VgivenZsingle}. 
\medskip

\noindent \textbf{M-step.} 
Here optimization of $Q(\btheta,\gamma)$ with respect ot $\btheta = (\brho, \bnu,\blambda)$ is performed under constraints~\eqref{eq:positiveConstraints},~\eqref{eq:rho-constraint}. Since $Q$ decomposes into a function of $(\brho,\bnu)$ and a function of $\blambda$, and since the constraints on $\brho,  \bnu$ and $\blambda$ are independent, maximization can be performed independently over the two blocks.  Indeed, gathering terms not depending on $\btheta$ into a constant $C$, 

\begin{multline*}
  Q(\btheta, \bgamma)  
    =\sum_{i=1}^n\Big[\sum_{k=1}^{K}\gamma_{i,k}\big[\log \pi_k  % \sum_{l\in\alpha_k}{\mb{\rho_k}}_l
  % -
    + \log \diri_k(\mb{W}_{i,k} |\brho_k ,\nu_k) % (|\alpha_k|+1)\log r_{i,k}
  % \\
  % &
      + \sum_{l\in\alpha_k^c}\log\dnoise( V_{i,l} | \lambda_k) % Exp_{\lambda_k}({\mb{v}_i}_l - 1)
      \big]\\
    + \sum_{k = K+1}^{K+d_1}\gamma_{i,k} \big[% \log\frac{1}{d} -2\log {\mb{v}_i}_j +
    \sum_{\ell\neq k}\log\dnoise( V_{i,l} | \lambda_k) %\log Exp_{\lambda_k}({V}_{i,\ell} - 1)
    \big] \Big]
    +C
   = Q_1(\brho, \bnu ) + Q_2(\blambda ) + C,
\end{multline*}
where 
\begin{align*}
  Q_1 (\brho, \bnu)& =\sum_{i=1}^n\sum_{k=1}^K\gamma_{i,k}\big[\log\sum_{l\in\alpha_k}{\mb{\rho_k}}_l + \log \diri_k(\mb{W}_{i,k} |\mb{\rho_k},\nu_k)\big] \,\\
  % Q_2(\blambda, \bgamma) & =\sum_{i=1}^n\big[\sum_{k=1}^K\gamma_{i,k}\sum_{l\in\alpha_k^c}\log \dnoise({V}_{i,l} |\lambda_k) %%Exp_{\lambda_k}({\mb{v}_i}_l - 1) +
  % \sum_{j = K+1}^{K+d_1}% \mathbb{E}}
  % \gamma_{i,e_j} \sum_{l\neq j}\log \dnoise({V}_{i,l} |\lambda_k) % Exp_{\lambda^{'}_j}({\mb{v}_i}_l - 1)
  % \big]
                             Q_2(\blambda) & =\sum_{i=1}^n\sum_{k=1}^{K+d_1}\gamma_{i,k}\sum_{l\in\alpha_k^c}\log \dnoise({V}_{i,l} |\lambda_k) \,.%%Exp_{\lambda_k}({\mb{v}_i}_l - 1) +
\end{align*}
Here we set  $\alpha_k = \{k-K\}$ for  $K<k\le K+d_1$, in accordance with the notations from Section~\ref{sec:mixtureModel}.
Notice that  the dependence of $Q_1$ and $Q_2$  on $\bgamma$ is omitted for the sake of concision. 
With these notations 

\begin{align*}
  \max_{ \substack{\btheta   \text{ s.t.}  \\ \eqref{eq:positiveConstraints},~\eqref{eq:rho-constraint}  }} Q(\btheta, \bgamma) =
  \max_{ \substack{\brho, \bnu \text{ s.t.} \\ \eqref{eq:rho-constraint}, \nu_{k}>0 , \;k\le K  }} Q_1(\brho, \bnu) +
  \max_{ \substack{ \blambda \text{ s.t.} \\\lambda_k>0, \; 1\le k\le K+d_1 }} Q_2(\blambda)   
\end{align*}
The function $Q_1$ being non-concave we use the python package
\textbf{mystic} (\cite{mckerns2012building}) to maximize it. 
For our choice of translated exponential noise,  $\dnoise(v|\lambda_k) = \lambda_k e^{-\lambda_k(v-1)} $, $v\ge 1$, the maximizer of $Q_2$ has an explicit expression, 

\begin{align*}
  \lambda_k^*=\frac{|\alpha_k^c|\sum_{i=1}^n\gamma_{i,k}}{\sum_{i=1}^n\gamma_{i,k}\sum_{l\in\alpha_k^c}( V_{i,\ell} - 1)}\;, \qquad k\le K+d_1.
\end{align*}
%%% Local Variables:
%%% mode: latex
%%% ispell-local-dictionary: "american"
%%% TeX-master: "Anomaly_VisualDisplay.tex"
%%% End:
\begin{remark}\label{rem:convEM}
  Let $\bgamma^t$ and $\btheta^t$ be the results of the $t$-th iteration of the algorithm then we conclude the iterative process if $Q(\btheta^t, \bgamma^t) < Q(\btheta^{t-1}, \bgamma^{t-1}) + \epsilon$, with $\epsilon$ a small threshold.
\end{remark}

%%% Local Variables:
%%% mode: latex
%%% ispell-local-dictionary: "american"
%%% TeX-master: "Anomaly_VisualDisplay_Long.tex"
%%% End:

\section{Graph-based Clustering and Visualization Tools}\label{sec:viz}
%\todo{ !! $n_0$ is the number of extreme data, not $n$}
Beyond the hard clustering that may be straightforwardly deduced from the computation of the likeliest values $z_1,\; \ldots,\; z_{n_0}$ for the hidden variables given the ${\mb V}_i$'s and the parameter estimates produced by the EM algorithm, % described in the Supplementary, % described in subsection \ref{subsec:inference}
 the statistical model previously introduced defines a natural structure of undirected weighted graph on the set of observed extremes, which interpretable layouts (graph drawing) can be directly derived % from,
 using classical solutions.
Indeed, a partition (hard clustering) of the set of (standardized) anomalies/extremes ${\mb V}_1,\; \ldots,\; {\mb V}_{n_0}$ is obtained by assigning membership of each ${\mb V}_i$ in a cluster (or cone/sub-simplex ) determined by the component of the estimated mixture model from which it arises with highest probability: precisely,
one then considers that the abnormal observation ${\mb V}_i$ is in the cluster indexed by

$$k_i=\argmax_{k\in\{1,\; \ldots,\; K+d_1 \}}\gamma_{i,k}$$
and is of type $\alpha_{k_i}$. However, our model-based approach brings much more information and the vector of posterior probabilities $(\gamma_{i,1},\; \ldots,\; \gamma_{i,K+d_1})$ output by the algorithm actually defines soft membership and represent the uncertainty in whether anomaly ${\mb V}_i$ is in a certain cluster. It additionally induces a similarity measure between the anomalies: the higher the probability that two extreme values arise from the same component of the mixture model, the more similar they are considered. Hence, consider the undirected graph whose vertices, indexed by $i=1,\; \ldots,\; n_0$, correspond to the extremal observations ${\mb V}_1,\; \ldots,\; {\mb V}_{n_0}$ and whose edgeweights are $w_{{\btheta}}({\mb V}_i,{\mb V}_j)$, $1\leq i\neq j\leq n_0$, where

$$
w_{{\btheta}}({\mb V}_i,{\mb V}_j)=\mathbb{P}\left(\mb Z_i=\mb Z_j \mid {{\mb V}}_i={\mb V}_i,\; {{\mb V}}_j={\mb V}_j,\; {\btheta} \right)=\sum_{k=1}^{K+d_1} \gamma_{i,k}\gamma_{j,k}.
$$
Based on this original graph description of the set of extremes, it is now possible to rank all anomalies (\textit{i.e.} extreme points) by degree of similarity to a given anomaly ${\mb V}_i$

$$
w_{{\btheta}}({\mb V}_i,{\mb V}_{(i,1)})\geq w_{{\btheta}}({\mb V}_i,{\mb V}_{(i,2)})\geq \ldots \geq w_{{\btheta}}({\mb V}_i,{\mb V}_{(i, n_0)})
$$
 and extract neighborhoods $\{ {\mb V}_{(i,1)},\; \ldots,\; {\mb V}_{(i,l)} \}$, $l\leq n_0$.
\medskip

\noindent {\bf Graph-theoretic clustering.} We point out that many alternative methods to that consisting in assigning to each any anomaly/extreme its likeliest component (\textit{i.e.} model-based clustering)  can be implemented in order to partition the similarity graph thus defined into subgraphs whose vertices correspond to  
similar anomalies, ranging from tree-based clustering procedures to techniques based on local connectivity properties through spectral clustering. One may refer to \textit{e.g.} \cite{SCHAEFFER200727} for an account of graph-theoretic clustering methods.
\smallskip

\noindent {\bf Graph visualization.} In possible combination with clustering, graph visualization techniques (see \textit{e.g.} \cite{WICS:WICS1343}), when the number $n_0$ of anomalies to be analyzed is large, can also be used to produce informative layouts. Discussing the merits and limitations of the wide variety of approaches documented in the literature in this purpose is beyond the scope of this paper. The usefulness of the weighted graph representation  proposed above combined with  state-of-the-art graph-mining tools is simply illustrated in 
Section~\ref{sec:flights_visu} and~\ref{sec:shuttle_visu}. We point out however that alternatives to the (force-based) graph drawing method used therein can be naturally considered, re-using for instance the eigenvectors of the graph Laplacian computed through a preliminary spectral clustering procedure (see \textit{e.g.} \cite{Athreya17} and the references therein for more details on spectral layout methods).

% The next section contains a simple illustration of

%%% Local Variables:
%%% mode: latex
%%% ispell-local-dictionary: "american"
%%% TeX-master: "Anomaly_VisualDisplay_Long.tex"
%%% End:

\section{Illustrative Experiments}\label{sec:experiment}

% It is the purpose of this section to display numerical results that provide strong empirical evidence of the relevance of the approach we promote for anomaly clustering/visualization. Favourable comparisons with state-of-the-art methods standing as natural competitors are also presented.
The aim of our experiments is double. First, investigate the goodness of fit of the Dirichlet mixture model fitted \emph{via} the EM algorithm on simulated data from the model. Second, provide empirical evidence of  the relevance of the approach we promote for anomaly clustering/visualization with real world data. Comparisons with state-of-the-art methods standing as natural competitors are  presented  for this purpose.

\subsection{Experiments on Simulated Data}\label{sec:simudata_clustering}
To assess the performance of the proposed estimator of the dependence structure and of the EM algorithm, we generate synthetic data according to Model~\ref{model:noisy}. 
%To try out the efficiency of the EM algorithm, we generate datasets straight from the mixture model developped in \ref{sec:model}.
The dimension is fixed to $d=100$ and the mixture components, that is the elements of  
$\mathbb{M}=\{\alpha_1,\ldots,\alpha_K\}$, are  randomly chosen in the power set of $\dd$  with $K=50$. The coefficients of the matrix  $\rho$ which determines the weights and centers through Eq. (14) in the Supplementary Material is also randomly chosen, then its columns are normalized so that  the moment constraint is satisfied. Finally. we fix $\nu_k=20$ for $1\leq k \leq K$ and  $\lambda_k$, $1\leq k \leq K+d_1$, are successively set to $1$, $0.75$, $0.5$, $0.25$ and $0.1$ to vary the noise level in the experiments. Then each point ${ \mb V}_i=R_i \mb{W}_i+ \bepsilon_i$, $i\leq n$, is generated with probability $\pi_k,k\in\{1,\ldots,K\}$ according to the mixture component $k\le K$, that is 

\begin{equation*}
  R_i \sim Pareto(1)|\{R_i>r_0\},\; 
  \mb{W}_i \sim \Phi_k,\; 
  \varepsilon_{i,j} \sim 1 + \dexp(\lambda_k),\; j\in\{1,\ldots,d\}\setminus\alpha_k,
\end{equation*}
and  with probability $\frac{1}{d}$ according to component  $k\in\{K,\ldots,K+d_1\}$ in such a way that

\begin{equation*}
  R_i \sim Pareto(1)|\{R_i>r_0\},\; 
  \mb{W}_i=1,\;
  \varepsilon_{i,j} \sim 1 + \dexp(\lambda_k),\; j\in\{1,\ldots,d\}\setminus\{k\}.
\end{equation*}
The threshold $r_0$ above which points are considered as extreme is fixed to $100$.
On this toy example, the pre-processing step that consists in applying DAMEX for recovering $\mathbb{M}$ produces an exact estimate, so that $\hat{\mathbb{M}} = \mathbb{M}$. Then the procedure described in Algorithm~\ref{algo:EM} is applied.
Tables \ref{error-1e3} and \ref{error-2e3} show the average absolute errors for the estimates $\widehat\rho$, $\widehat\nu$ and $\widehat\lambda$ on $50$ datasets of the $n_0$ generated extreme points, for $n_0 = 1e+3, 2e+3$,  namely 

\begin{align*}
  \err(\widehat\rho)&=\frac{1}{50\cdot K\cdot d}\sum_{l=1}^{50}\sum_{k=1}^K\sum_{j=1}^d|\widehat\rho_{k,j} - \rho_{k,j}|\\
  \err(\widehat\nu)&=\frac{1}{50\cdot K}\sum_{l=1}^{50}\sum_{k=1}^K|\widehat\nu_k - \nu_k|\\
  \err(\widehat\lambda)&=\frac{1}{50\cdot (K+d_1)}\sum_{l=1}^{50}\sum_{k=1}^{K+d_1}|\widehat\lambda_k - \lambda_k|
\end{align*}
On this toy example, estimates of the means and weights, as well as those of the noise parameters, are  almost exact. In contrast, the estimates of the  $\nu_k$'s are not that accurate, but, as shown next, this drawback does not jeopardize cluster identification. 

\begin{table}[!ht]
  \caption{Average error on the model parameters, $n_0=1e3$ extreme points}
  \begin{center}
    \label{error-1e3}
    \begin{tabular}{|l|c|c|c|c|c|}
      \hline
      &  $\lambda_k=1.$ & $\lambda_k=0.75$ & $\lambda_k=0.5$ & $\lambda_k=0.25$ & $\lambda_k=0.1$ \\
      \hline
      $err(\widehat\rho)$ & 1.39e-5 & 1.37e-5 & 1.57e-5 & 1.22e-5 & 2.11e-5 \\
      \hline
      $err(\widehat\nu)$ & 5.53 & 5.81 & 6.28 & 6.41 & 9.06 \\
      \hline
      $err(\widehat\lambda)$ & 2.65e-2 & 2.04e-2 & 1.19e-2 & 5.97e-3 & 3.66e-3 \\
      % \hline
      % &  $\lambda_k=1.$ & $\lambda_k=0.5$ & $\lambda_k=0.1$ \\
      % \hline
      % $err(\widehat\rho)$ & 1.39e-5 & 1.57e-5 &  2.11e-5 \\
      % \hline
      % $err(\widehat\nu)$ & 5.53 & 6.28 &  9.06 \\
      % \hline
      % $err(\widehat\lambda)$ & 2.65e-2 & 1.19e-2  & 3.66e-3 \\
      \hline
    \end{tabular}
  \end{center}
\end{table}

\begin{figure}[ht]
  % \centerline{\includegraphics[width=1.1\linewidth, scale=0.18, left]{figures/test_20_10_100pts_3}}
  \centerline{\includegraphics[width=0.6\linewidth]{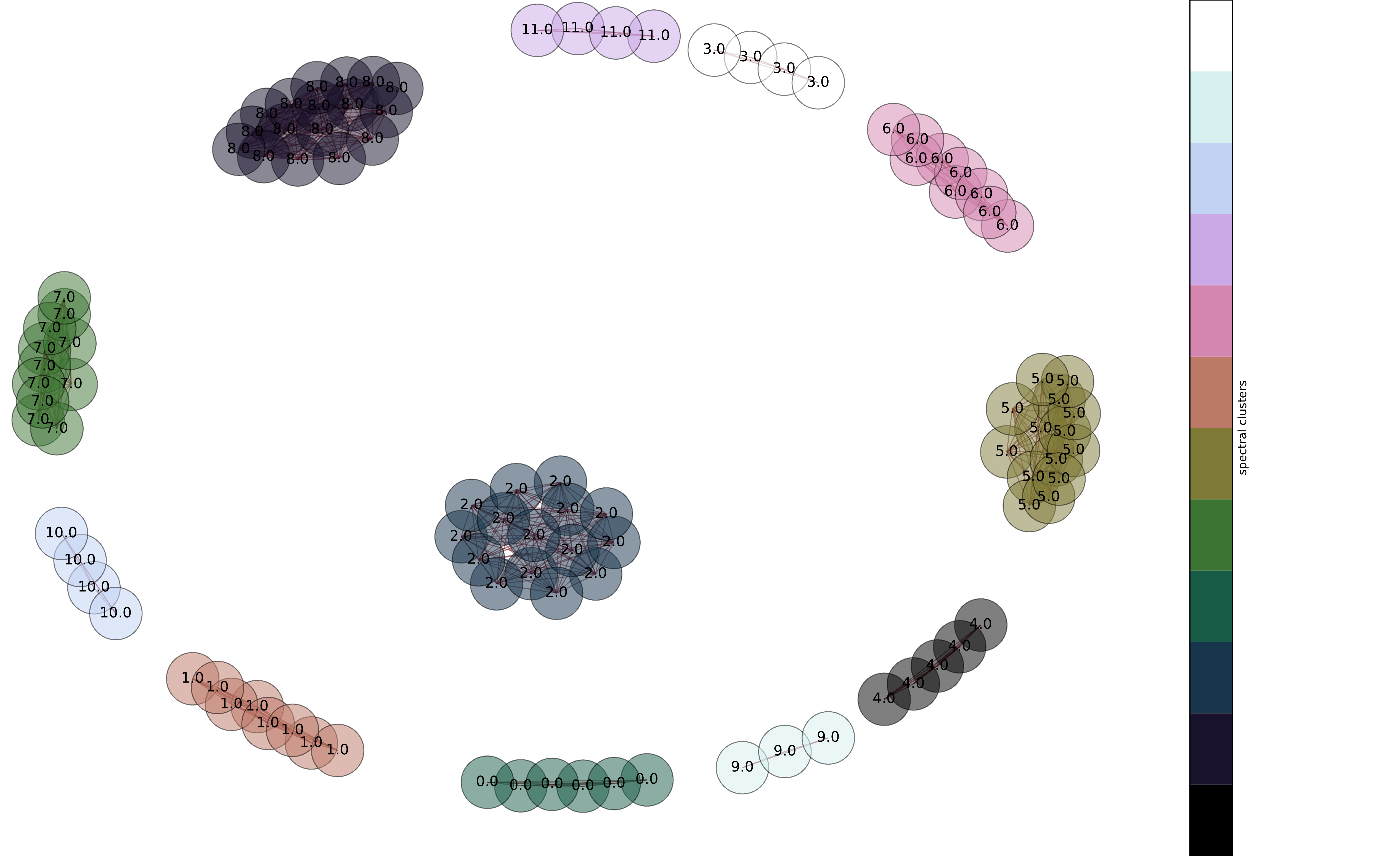}}%}{figures/test_20_10_100pts_3}}
  \caption{Spectral clustering visualization of  a synthetic anomaly test data of size
    $100$ with $d=20$ and $|\mathbb{M}|=12$. \newline  Each point is represented as a
  numbered node. The numbers indicate the true labels, while the colors 
  correspond to the clusters produced by the spectral clustering method. The spatial arrangement of the nodes is
  obtained by the Fr\"uchtermen-Reingold
  algorithm.} \label{fig:toy-example}   % and numbered nod represents a new point, which label is indicated by the number number on each node represents its true label.

\end{figure}

\begin{table}[!ht]
  \begin{center}
    \caption{Average error on the model parameters, $n_0=2e3$ extreme points}%%  \caption{$n_0=2e3$}
    \label{error-2e3}
    \begin{tabular}{|l|c|c|c|c|c|}
      \hline
      &  $\lambda_k=1.$ & $\lambda_k=0.75$ & $\lambda_k=0.5$ & $\lambda_k=0.25$ & $\lambda_k=0.1$ \\
      \hline
      $err(\widehat\rho)$ & 9.98e-6 & 1.12e-5 & 1.06e-5 & 1.62e-5 & 1.64e-5 \\
      \hline
      $err(\widehat\nu)$ & 3.23 & 4.13 & 4.08 & 4.29 & 5.05 \\
      \hline
      $err(\widehat\lambda)$ & 1.62e-2 & 1.2e-2 & 8.11e-3 & 4.28e-3 & 3.11e-3 \\
      \hline
    \end{tabular}
    \end{center}
\end{table}
The performance in terms of cluster identification is measured as follows: for each point ${ \bm v}_i$, the true label $y_i\in\{1,\ldots,K+d_1\}$ is compared with the label obtained \emph{via} assignment to the highest probable component, that is $\hat y_i =  \argmax_{k\in\{1,\; \ldots,\; K+d_1 \}} \gamma_{i,k}$. 
Table~\ref{error-labels} shows the average number of labeling errors for different values of $n_0$ and $\lambda_k$.  % by comparing the true label $y_i\in\{1,\ldots,K+d_1\}$ and $\argmax_{k\in\{1,\; \ldots,\; K+d_1 \}} \gamma_{i,k}$.
\begin{table}[!ht]
  \begin{center}

          \caption{Average number of labeling errors}    \label{error-labels}
    \begin{tabular}{|l|c|c|c|c|c|}
      \hline
      &  $\lambda_k=1.$ & $\lambda_k=0.75$ & $\lambda_k=0.5$ & $\lambda_k=0.25$ & $\lambda_k=0.1$ \\
      \hline
      $n_0=1e3$ & 0. & 0. &  0. & 0.6 & 264.4 \\
      \hline
      $n_0=2e3$ & 0. & 0. &  0.4 & 1.8 & 537.8 \\
      \hline
    \end{tabular}
  \end{center}
%%  \caption*{Average number of label's errors in fonction of $\lambda$ and $n_0$}
\end{table}
% \hline
%       &  $\lambda_k=1.$  & $\lambda_k=0.5$ & $\lambda_k=0.25$ & $\lambda_k=0.15$ & $\lambda_k=0.1$ \\
%       \hline
%       $n_0=1e3$ & 0. & 0. & 0.4 & 70.0 & 264.4 \\
%       \hline
%       $n_0=2e3$ & 0. &  0. & & & 537.8 \\
%       \hline
Figure~\ref{fig:toy-example} illustrates the relevance  of the proposed
approach regarding anomaly visualization. A test set of size $100$  consisting of extreme data 
% $100$ new datapoint
is simulated as above, and the corresponding matrix $\widehat\bgamma$
is computed according to~\eqref{eq:gamma-ik} % in the Supplementary Material
with ${\btheta}$ taken as
the output of the training step (\ie\ Algorithm~\ref{algo:EM} run with
the training dataset of $n_0 = 2e3$ points). Finally an adjacency
matrix $w_{\widehat{\btheta} }({\mb v}_i,{\mb v}_j)$ is
obtained as detailed in Section~\ref{sec:viz}, on which we apply spectral clustering
in order to group the points according to the similarities measured by $w$.
Graph visualization of $w$ is next performed % of the clusters of 'extreme flights' is made
using the python package 'Networkx' \cite{hagberg-2008-exploring},
that provides a spring layout of the graph according to the Fr\"uchtermen-Reingold algorithm, see \cite{fruchterman1991graph}.
A hard thresholding is applied to the edges in $w$ in order to improve  readability:  % done before the visualisation to cut
edges $(i, j)$
such that $w_{\widehat{\btheta} }({\mb v}_i,{\mb v}_j) < \epsilon$ with $\epsilon$ a small threshold are removed. 
% to get a better view of the graph.
Each cluster output by the spectral clustering method is identified  with a specific color.
% 
% % The output of the spectral clustering on a dataset $({v}_1,\ldots,{v}_{100})$ obtained by generating $100$ new points.
% The adjacency matrix given by $w_{\widehat{\btheta} % , \widehat\blambda
% }({\mb v}_i,{\mb v}_j)$ was computed after estimating $\widehat{\btheta}$ % and $\widehat\blambda$
% on a training dataset of $2000$ generated points.

\subsection{Flights Clustering and Visualization}\label{sec:flights_visu}
\begin{figure}[ht]
  % \centerline{\includegraphics[width=1.1\linewidth, scale=0.18, left]{figures/dmx_agglo_1000_05_300pts}}
    \centerline{\includegraphics[width=0.6\linewidth]{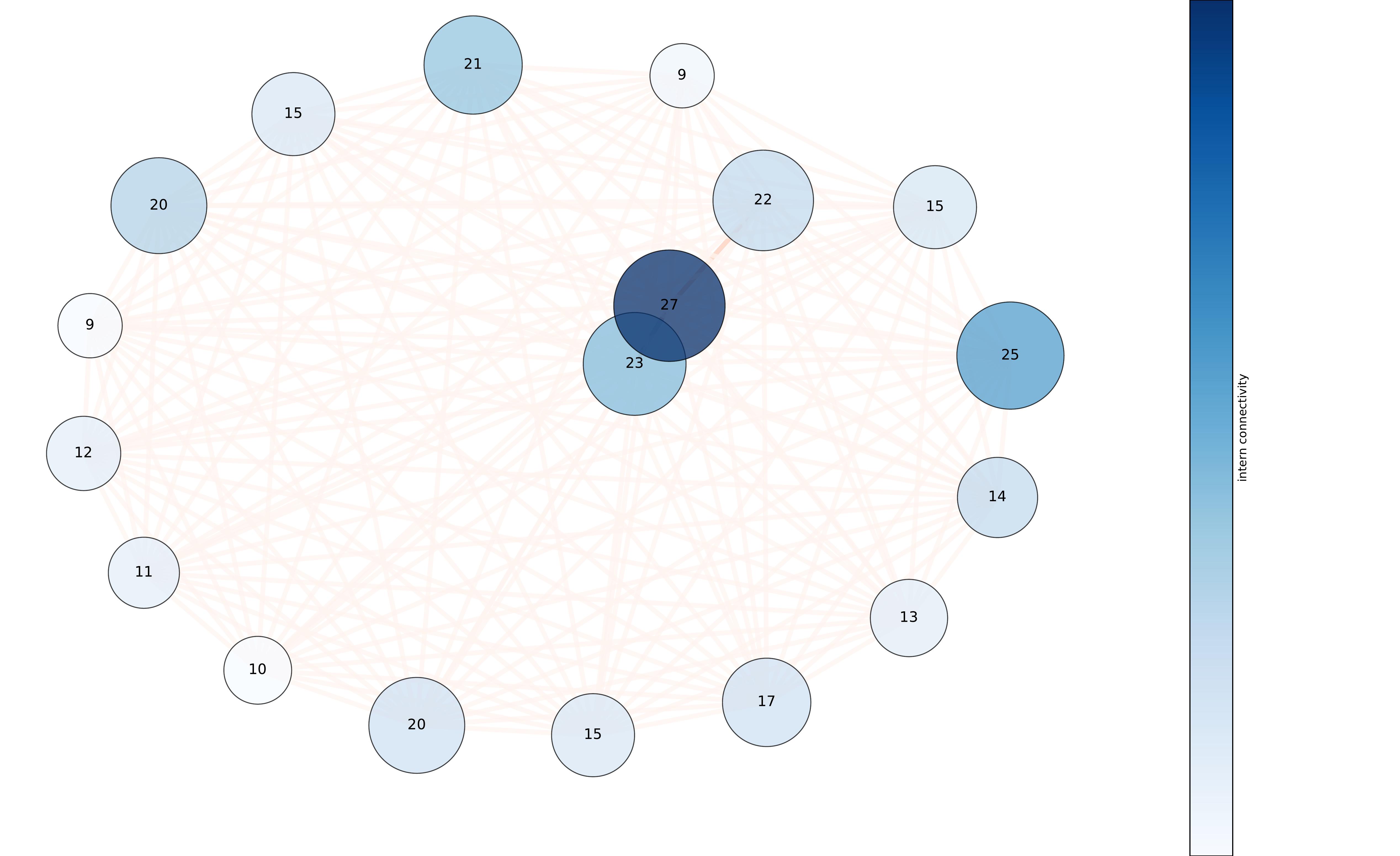}}%figures/dmx_agglo_1000_05_300pts}}
  \caption{Spectral clustering visualization  
    of flights anomalies with agglomerated nodes. \newline
  The agglomerated visualization is obtained
  \emph{via} spectral clustering: each node represents a
  cluster.  Levels of blue show the intern connectivity between the
  original nodes so that darker clusters have strongly connected
  elements. The size of each node is proportional to  the number of points forming the cluster. } \label{fig:agglo-flights} 
\end{figure}

The methodology proposed is currently tested by Airbus to assist in  building  health indicators for condition based maintenance.
Health indicators are  used for assessing the current state of some system and also for forecasting its future states and possible degradation (\emph{e.g.} bleed, power systems, engine, APU,~\dots ).
Airlines can be then informed that some systems should be maintained, so as to avoid any operational procedure at a given time horizon susceptible to cause \textit{e.g.}  delays, operational interruptions, \emph{etc}~\dots.
The construction of a health indicator can be basically summarized as follows:
\begin{enumerate}
\item Collect health and usage data from various aircrafts (generally one has to consider similar ones).
\item Collect some operational events happening on these aircrafts due to some aircraft system errors (\textit{e.g.} operational interruption, delays)
\item Identify anomalies in the health and usage data.
\item Identify some dependencies between health and usage data anomalies and operational events (by means of statistical hypothesis testing but also thanks to human expertise).
\item As soon as some dependencies are well identified, a health indicator is built.
\end{enumerate}
The main barrier is the identification and the understanding of the anomalies. Different operational events are often recorded, corresponding to the degradation of different systems.
Usually, a first stage of anomaly detection is performed, followed by a clustering of the anomalies listed for interpretation purpose. The major advantage of the approach proposed in this paper is that it directly provides a similarity measure between the anomalies.
This strategy is illustrated by Fig.~\ref{fig:agglo-flights}. 
The  proposed method  was applied on a dataset of $18553$ flights, each of which is characterized by $82$ parameters. In order to differentiate between anomalies corresponding to unusually large and small values, each feature is duplicated and each copy of a given feature is defined as the positive (\emph{resp.} negative) value of the parameter above (\emph{resp.} below) its mean value.

\begin{figure}[ht]
  % \centerline{\includegraphics[width=1.1\linewidth, scale=0.18, left]{figures/dmx_1000_05_300pts}}
  \centerline{\includegraphics[width=0.6\linewidth]{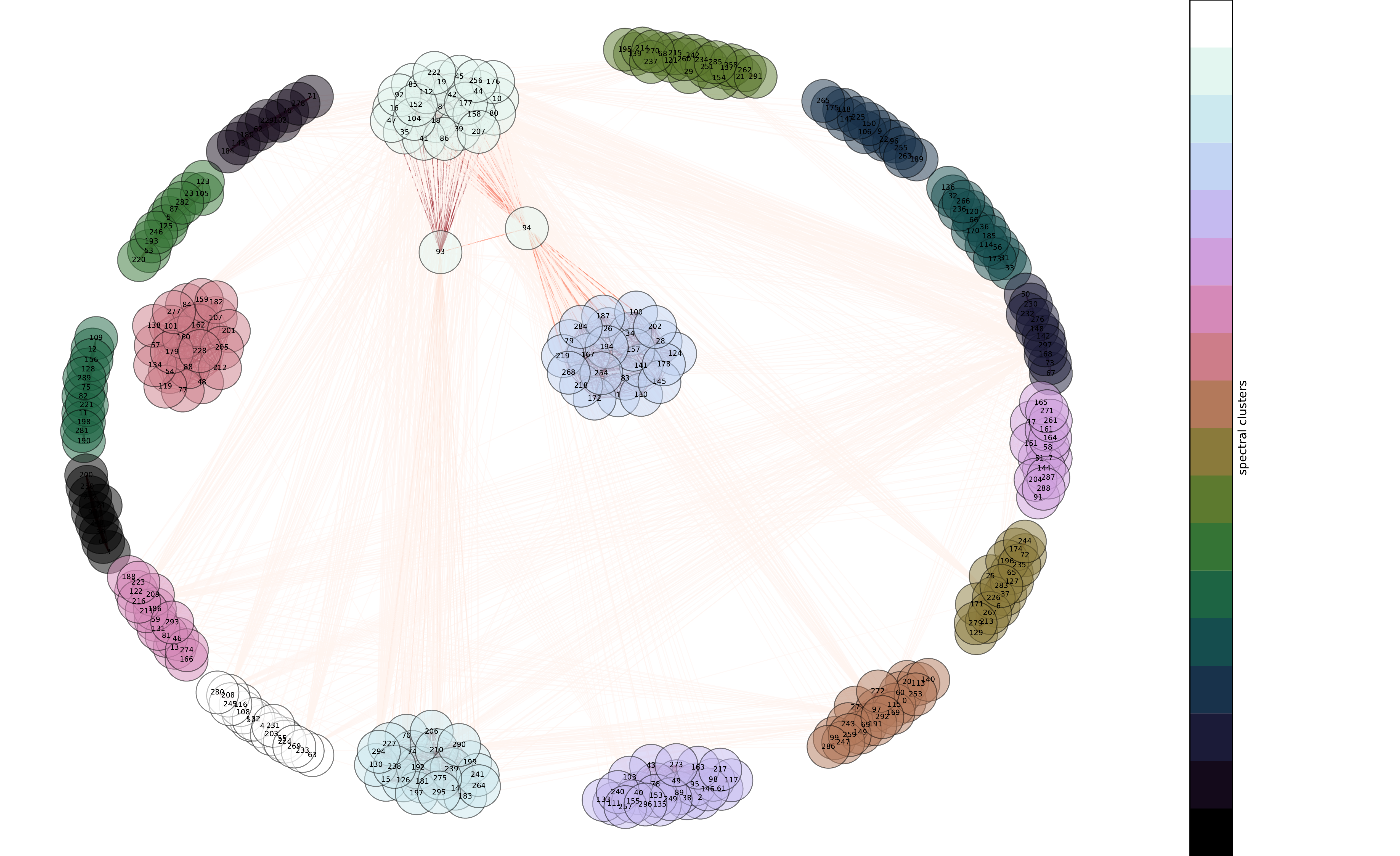}}%{figures/dmx_1000_05_300pts}}
  \caption{Spectral clustering visualization  
    of flights anomalies. \newline
    The number of each node is the (anonymized) flight identification number.  The nodes  colors  and the spatial arrangement are obtained similarly to Fig.~\ref{fig:toy-example}.
}\label{fig:numbered-flights}
%  \centerline{\includegraphics[scale=0.17]{figures/clf_1e3_03_500pts_30clst}}
  % \caption{Spectral clustering visualization  
  %   of flights anomalies}
  % Spectral clustering with agglomerated flights. The size of the nodes is proportional to the number of elements in it while levels of blues show the intern connectivity between the original nodes
  %   so that darker clusters have strongly connected elements.}
  % \label{fig:agglo-flights}
\end{figure}

% In order to take into account the higher and smaller values of each parameter, we duplicate 
% each feature into one that takes values above the mean and the other the values below the mean.
% The metodology developed along the article then gives us an estimate $\widehat{w}$ of the adjacency matrix on which we apply
% the spectral clustering algorithm.
% The visualisation of the clusters of 'extreme flights' is made using the python package 'Networkx' \cite{hagberg-2008-exploring},
% that displays the nodes according to the Fruchtermen-Reingold algorithm \cite{fruchterman1991graph}.
% Each node represents a cluster of flights, edges are added relatively to the number of connected components
% between the clusters and the size of the nodes is proportional to the number of points in the corresponding cluster.  
Fig.~\ref{fig:agglo-flights} and Fig.~\ref{fig:numbered-flights} display the clustering of $300$ 'extremal' flights into $18$ groups,  showing on the one hand the output of the spectral clustering applied to the similarity graph $w_{\hat {\btheta} }$  and on the other hand the underlying graph obtained with the same procedure as in  Fig.~\ref{fig:toy-example}. 
% <<<<<<< HEAD
%   \centerline{\includegraphics[scale=0.17]{figures/clf_1e3_03_500pts}}
%   \caption{Spectral clustering visualization  
%     of flights anomalies}\label{fig:numbered-flights}
%   The number of each node is the (anonymized) flight identification number.  The color scale and the spatial arrangement are obtained similarly to Figure~\ref{fig:toy-example}.
% =======

%%% Local Variables:
%%% mode: latex
%%% ispell-local-dictionary: "american"
%%% TeX-master: "Anomaly_VisualDisplay.tex"
%%% End:

\subsection{A Real World Data Experiment with the Ground Truth}\label{sec:shuttle_visu}
The \emph{shuttle} dataset is available in the UCI repository, see \cite{Dua:2017} (training and test datasets are merged here), $9$ numerical attributes and $7$ classes
are observed.
Class $1$ representing more than $80\%$ of the dataset, since our goal is to cluster rare and extreme events, instances from all classes but $1$ are analyzed, leading to a sample size equal to $12414$. The number of extreme points considered is denoted by $n_{0}$ here.
We compare our approach to the $K$-means algorithm
and the \emph{spectral clustering} algorithm as implemented in \cite{scikit-learn}. The number of clusters that we fix in advance to run each of these algorithms is denoted by $n_{cluster}$. The performance of each approach is evaluated by computing the \emph{purity} score:

\begin{align*}
  \text{\emph{purity}}=\frac{1}{n_{0}}\sum_{i=1}^{n_{cluster}}\max_{c\in C}n_{i,c},
\end{align*}
where $n_{i,c}$ is the number of elements of class $c$ in the cluster $i$.
As shown by Table \ref{tab:purity}, the purity score produced by the anomaly clustering technique promoted in this paper is always equal to or higher than those obtained by means of the other algorithms.

\begin{table}[!ht]
  \caption{Purity score - Comparisons with standard approaches for different extreme sample sizes.}
  \begin{center}
    \label{tab:purity}
    \begin{tabular}{|l|c|c|c|c|c|}
      \hline
      &  $n_{0}=500$ & $n_{0}=400$ & $n_{0}=300$ & $n_{0}=200$ & $n_{0}=100$ \\
      \hline
      Dirichlet mixture & 0.8 & 0.82 & 0.82 & 0.84 & 0.85 \\
      \hline
      Kmeans & 0.72 & 0.73 & 0.75 & 0.78 & 0.8 \\
      \hline
      % DBSCAN & 0.742 & 0.728 & 0.59 \\
      Spectral clustering & 0.78 & 0.77 & 0.82 & 0.81 & 0.8 \\
      \hline
    \end{tabular}
  \end{center}
\end{table}

%%% Local Variables:
%%% mode: latex
%%% ispell-local-dictionary: "american"
%%% TeX-master: "Anomaly_VisualDisplay_Long.tex"
%%% End:

\section{Conclusion}\label{sec:concl} 
Because extreme values (viewed as anomalies here) cannot be summarized
by simple meaningful summary statistics such as local means or
modes/centroids, clustering and dimensionality reduction techniques
for such abnormal observations must be of very different nature than
those developed for analyzing data lying in high probability
regions. This paper is a first attempt to design a methodology fully
dedicated to the clustering and visualization of anomalies, by means
of a statistical mixture model for multivariate extremes that can be
interpreted as a noisy version of the angular measure, which
distribution on the unit sphere exhaustively describes the limit
dependence structure of the extremes.
Mixture component are identified here with specific sub-simplices forming the support of the angular measure. 
% Localization of the mixture components is understood
% here as closeness of the data arising from them to a specific
% sub-simplex forming the support of the angular measure.
Considering
synthetic and real datasets, we also provide empirical evidence of
the usefulness of (graph-based) techniques that can be
straightforwardly implemented from the framework we developed.

\section*{Acknowledgements}
This work was supported by a public grant as part of the
Investissement d'avenir project, reference ANR-11-LABX-0056-LMH,
LabEx LMH.

\bibliographystyle{apalike}
\bibliography{mvextrem}
%\input{Supplementary}
%\input{Appendix}
% \section{Section Title}
% Main contents here.

% \subsection{Subsection Title}
% A figure in Fig.~\ref{fig:spiral}. Please use high quality graphics for your camera-ready submission -- if you can use a vector graphics format such as \texttt{.eps} or \texttt{.pdf}.
% \begin{figure}[htp]
% \begin{center}
% \includegraphics[width=0.5\textwidth]{spiral.eps}
% \caption{A spiral.}\label{fig:spiral}
% \end{center}
% \end{figure}

% An example of citation~\cite{DBLP:conf/acml/2009}.

% \acks{Acknowledgements should go at the end, before appendices and references.}

% %\bibliographystyle{plain}
% \bibliography{acml18}

% \appendix

% \section{First Appendix}\label{apd:first}

% This is the first appendix.

% \section{Second Appendix}\label{apd:second}

% This is the second appendix.

\end{document}